\documentclass[%
 reprint,
superscriptaddress,
 amsmath,amssymb,
 aps,
]{revtex4-2}

\usepackage{graphicx}
\usepackage{dcolumn}
\usepackage{bm}

\usepackage{physics}
\usepackage{cleveref}
\usepackage{mathrsfs} 
\usepackage{booktabs}
\usepackage{multirow}
\usepackage{makecell}
\usepackage[caption=false]{subfig} 
\usepackage{longtable}
\usepackage{upgreek}
\usepackage{xcolor}
\usepackage{float}

\newcolumntype{s}{>{\centering\arraybackslash}X}
\newcommand{\B}{\text{B}}
\newcommand{\p}{\text{p}}
\newcommand{\aerror}[2]{\substack{\scriptscriptstyle +#1\\ \scriptscriptstyle-#2}}

\begin{document}

\preprint{ADP-23-27/T1236}

\title{Study of the pion-mass dependence of $\rho$-meson properties in lattice QCD}

\author{Kang Yu}
\affiliation{%
School of Physical Sciences, University of Chinese Academy of Sciences, Beijing 100049, China
}%
\author{Yan Li}
\affiliation{
Department of Physics, University of Cyprus, 20537 Nicosia, Cyprus
}%
\author{Jia-Jun Wu}
\affiliation{%
School of Physical Sciences, University of Chinese Academy of Sciences, Beijing 100049, China
}%
\affiliation{%
Southern Center for Nuclear-Science Theory (SCNT), Institute of Modern Physics,
Chinese Academy of Sciences, Huizhou 516000, Guangdong Province, China
}%
\author{Derek B. Leinweber}
\affiliation{Special Research Centre for the Subatomic Structure of Matter (CSSM), Department
of Physics, University of Adelaide, Adelaide, South Australia 5005, Australia}
\author{Anthony W. Thomas}
\affiliation{Special Research Centre for the Subatomic Structure of Matter (CSSM), Department
of Physics, University of Adelaide, Adelaide, South Australia 5005, Australia}

\date{\today}

\begin{abstract}
We collect spectra extracted in the $I=\ell=1$ $\pi\pi$ sector provided by various lattice QCD collaborations and study the $m_\pi$ dependence of $\rho$-meson properties using Hamiltonian Effective Field Theory (HEFT).
In this unified analysis, the coupling constant and cutoff mass, characterizing the $\rho - \pi \pi$ vertex, are both found to be weakly dependent on $m_\pi$, while the mass of the bare $\rho$, associated with a simple quark-model state, shows a linear dependence on $m_\pi^2$.
Both the lattice results and experimental data can be  described well.
Drawing on HEFT's ability to describe the pion mass dependence of resonances in a single formalism, we map the dependence of the phase shift as a function of $m_\pi$, and expose interesting discrepancies in contemporary lattice QCD results.

\end{abstract}

\maketitle


\section{introduction}

One of the most significant challenges in hadron physics is to understand the internal composition of diverse hadrons.
However, because of the non-perturbative nature of the strong interaction in the low energy regime, the structure of hadrons within Quantum Chromodynamics (QCD) has remained unsolvable analytically. 
In order to develop insight into hadron structure and guide experimental work, a wide variety of phenomenological models have been developed. 
This includes the constituent quark model~\cite{Capstick:2000qj}, the MIT~\cite{DeGrand:1975cf} 
and cloudy bag models~\cite{Theberge:1980ye,Thomas:1981vc} as well others based upon Schwinger-Dyson equations~\cite{Roberts:1994dr} and molecular~\cite{Guo:2017jvc} and hybrid~\cite{Meyer:2015eta} models.
As it is often possible to adjust the parameters in these models in order to reproduce the limited experimental data, these models can typically not be distinguished solely on the basis of how well they describe experiments.

On the other hand, the approximate chiral symmetry of QCD means that the pion is a pseudo-Goldstone boson, with a much smaller mass, $m_\pi$, than other hadrons. 
Because the mass of the pion squared is proportional to the quark mass over a wide range, 
it is reasonable to expand certain physical variables in terms of $m_\pi$.
For instance, the mass of a resonance $R$ can be expressed perturbatively as follows:
\begin{align}
   m_R = \sum_{n=0}^{\infty} \alpha_n(\{g_i\})\, m_\pi^{2n} + \Sigma_R(\{g_i\},m_\pi), 
\label{eq:massexp}
\end{align}
where $\{g_i\}$ is the set of free parameters of the model and $\Sigma_R$ is the self-energy term. 
 The separation described in Eq.~\ref{eq:massexp} differentiates between the known model-independent coefficients of terms nonanalytic in the quark mass which are contained within the self-energy terms, and the unknown coefficients of terms analytic in the quark mass, coefficients that are constrained by fitting data.

In the real world, $m_\pi$ takes the fixed value $\mu_\pi\approx 138.5$ MeV (for charged pions), and it is typically feasible to adjust the free parameters of a model, $\{g_i\}$, to reproduce the correct experimental value, $m_R(\mu_\pi)$.
However, when we extend the model to unphysical pion masses these models may predict different values of $m_R$ versus $m_\pi$. 
The dependence of various physical variables on $m_\pi$ offers a fresh perspective in exploring the structure of hadrons in the non-perturbative regime~\cite{Thomas:2002sj}. 
As a result, it is of great significance to make measurements on $m_R$ at unphysical $m_\pi$ values.

Lattice QCD (LQCD) is a well-established non-perturbative formulation of QCD, defined on a finite and discretized volume of four-dimensional Euclidean space-time.
Through simulation of the two-point Green functions of composite operators, one can obtain the finite volume spectrum of eigenvalues of the QCD Hamiltonian, with specific quantum numbers, as a function of $m_\pi$. 
We stress that such results are genuine predictions of QCD, even though the light quark masses do not take their physical values.
Moreover, the phase shift in the infinite volume can then be obtained through the well-known L\"{u}scher formula and its extensions~\cite{Luscher:1990ux,Kim:2005gf,Hansen:2012tf,He:2005ey,Rummukainen:1995vs,Gockeler:2012yj}.

LQCD has developed rapidly over the half century since Wilson's pioneering work was published in 1974~\cite{Wilson:1974sk}.
With the substantial progress in simulation algorithms and tremendous advances in computing power, many LQCD collaborations~\cite{FlavourLatticeAveragingGroupFLAG:2021npn, Briceno:2017max} have extracted the finite volume spectra for various sets of quantum numbers, including the $\rho$ meson, over a range of values of $m_\pi$.

Experimentally, the $\rho$ meson is identified as a broad peak around $\sqrt{s}=770$ MeV in the invariant mass distribution of the isovector $P$-wave of $\pi\pi$ 
scattering~\cite{ParticleDataGroup:2022pth}. It is often identified as a confined $q\bar{q}$ state, consistent with the constituent quark model.
This picture is supported by several theoretical arguments, such as the the large-$N_c$ limit of 
QCD~\cite{Aceti:2012dd,Jaffe:2007id,Pelaez:2003dy,Pelaez:2006nj}. 
Nevertheless, the sizable decay width, $\Gamma_{\rho\to\pi\pi}\approx 140$ MeV~\cite{ParticleDataGroup:2022pth}, signifies the $\rho$ meson's strong coupling to the $\pi\pi$ channel. 
In other words, the observed peak structure results from the interaction between a $q\bar{q}$ state, referred to as the bare $\rho$, and the $\pi\pi$ continuum at the hadronic level.
Consequently, a comprehensive study of the $\rho$ meson necessitates an exploration of the $\pi\pi$ scattering sector.

In the past decade, multiple LQCD groups have provided energy levels for the $P$-wave $\pi\pi$ sector~\cite{Lang:2011mn, CS:2011vqf, Bali:2015gji, Wilson:2015dqa, Dudek:2012xn, ExtendedTwistedMass:2019omo, Alexandrou:2017mpi, Pelissier:2012pi, Feng:2010es, Guo:2016zos, Fu:2016itp, Andersen:2018mau,Fischer:2020yvw,  Rodas:2023gma}.
However, there has been little work collating spectra from various collaborations~\cite{Hu:2017wli, Hu:2016shf}, particularly for $N_f=2+1$, and performing a consistent unified analysis. 
That is the aim of this paper.

The L\"{u}scher formula is the most practical way to relate lattice calculations to the elastic scattering phase shifts of two spinless particles.
Therefore, when dealing with a system containing only one $\pi\pi$ channel, it is sufficient to utilize the standard L\"{u}scher formula to relate the finite volume spectrum to the phase shifts.
However, in our present study, we also aim to incorporate the $\omega\pi$ channel, in order to assess its impact. 
While it is not an open channel, it does generate the leading non-analytic behaviour of the $\rho$ mass as a function of $m_\pi$.
%

While the L\"uscher formalism can certainly include the additional
$\omega\pi$ channel, we note that there are other approaches which provide computational
convenience with very little overhead in incorporating several
two-particle channels. 
As an example, we mention that the unitarized chiral perturbation theory ($\chi$PT) can calculate the finite volume spectrum through the pole position of the T matrix defined in the finite volume as shown in Refs.~\cite{Doring:2011ip, Doring:2012eu}.

Alternatively, Hamiltonian effective field theory (HEFT) also incorporates the L\"{u}scher formalism and establishes a connection between the scattering process in infinite volume and the finite volume spectrum of the system~\cite{Wu:2014vma}. 
For multi-channel scattering, the advantages and practicality of HEFT have been demonstrated in studies of various resonances, including the Roper~\cite{Liu:2016uzk,Wu:2017qve}, the $\Lambda(1405)$ ~\cite{Liu:2016wxq}, the $N^*(1535)$~\cite{Liu:2015ktc} and the $D^*_s(2317/2460)$~\cite{Yang:2021tvc}.
Because the Hamiltonian operates within the Fock space, effectively describing interactions among various different channels, the HEFT approach has two important features. 
1) It provides insight into the composition of the eigenstates through the strength of various components of the eigenvectors. 2) It also enables an examination of the quark-mass dependence of resonance properties in a single formalism, enabling this unified analysis.

Here, we consistently analyze the spectra provided by several different LQCD collaborations using the HEFT framework, drawing on results extracted in the rest frame~\cite{Li:2019qvh}, moving frame and elongated frames~\cite{Li:2021mob}.
Motivated by the physical picture mentioned, the Hamiltonian employed here is studied within a framework that involves a bare $\rho$, as well as $\pi\pi$ 
and $\omega\pi$ channels.
We obtain the bare $\rho$ mass in various regularization schemes from the lattice energy levels and investigate its dependence on $m_\pi$. 
We observe that the linear slope of the bare $\rho$ mass with respect to $m_\pi^2$ is minimally affected by scheme dependence.
Furthermore, we investigate the composition of the $\rho$ meson, using the eigenvector of the eigen-energy state closest to the physical $\rho$ mass.
Finally, we illustrate interesting discrepancies in contemporary lattice QCD calculations.

The paper is organized as follows. 
In Sec.~\ref{sec:formalism} we provide an overview of the HEFT formulation and proceed to construct the finite volume Hamiltonian for the specific case under investigation. 
Section~\ref{sec:discussion} presents the results of the numerical analysis and examines the dependence of various variables on $m_\pi$. 
Finally, in Sec.~\ref{sec:summary}, we draw the discussion to a close with a concise summary and a suggestions for further analysis.  

\section{Formalism}
\label{sec:formalism}

\subsection{Hamiltonian model}
The Hamiltonian in the center of mass frame of the interacting system is divided into two parts as follows,
\begin{align}
   H = H_0 + H_I \, , 
\label{eq:H}
\end{align}
where $H_0$ is the non-interacting part, and $H_I$ is the interaction part.
In this work, we include a bare $\rho$ meson, which can be identified as a $q\bar{q}$ state, as well as two coupled channels, $\pi\pi$ and $\pi\omega$.
In the infinite volume, characterized by $\mathrm{SO}(3)$ symmetry, it is most convenient to express the interaction in the $JLS$ basis defined as~\cite{Weinberg:1995mt}
\begin{align}
   \ket{\alpha;k^*,JM\ell S} = &A_\alpha
   \sum\limits_{m\,\sigma\,\sigma_1\,\sigma_2}C_{\ell S}(JM;m\sigma)\,C_{s_1s_2}(S\sigma;\sigma_1\sigma_2) \notag \\
   &\times\int d\hat{\boldsymbol{k}}^*\,Y_{\ell m}(\hat{\boldsymbol{k}}^*)\ket{\alpha;\boldsymbol{k}^*,\sigma_1\,\sigma_2} \, , 
\label{eq:inf JMLS basis}
\end{align}
where $C_{j_1j_2}(jm;m_1m_2)$ is the Clebsch-Gordon (CG) coefficient of the $\mathrm{SU}(2)$ group, $Y_{lm}$ are the normalized spherical harmonics functions and 
$|\,\alpha;\boldsymbol{k}^*,\,\sigma_1\,\sigma_2\rangle$ indicates the $\alpha=\pi\pi$ or $\pi\omega$ channels, with relative momentum $\boldsymbol{k}^*$ and $z$-components of the spins of two particles, $\sigma_1$ and $\sigma_2$, respectively.
(For convenience, the quantities with an asterisk in this paper are all defined in the center of mass frame.)    
In addition, $A_\alpha$ is the normalization factor that equals $\frac{1}{\sqrt{2}}$ if $\alpha=\pi\pi$ and otherwise is unity, $m$ and $\sigma$ are the z-components of the orbital angular momentum and total spin, respectively, 
$S_1$ and $S_2$ are the spins of the particles in channel $\alpha$ and $J$, $M$, $\ell$ and $S$ are the total angular momentum, the z-component of the total angular momentum, orbital angular momentum, and total spin, respectively.
The $\ket{\alpha;k^*,M}$ is normalized as,
\begin{align}
   \braket{\alpha;k^*,JM\ell S}{\alpha';k^{*\,\prime},JM^\prime\ell S} &= \frac{\delta(k^*-k^{*\,\prime})}{k^{*\,2}}\,\delta_{\alpha\alpha^\prime}\,\delta_{MM^\prime} \, . 
\label{eq:normal:JLS:inf}
\end{align}

In general, there are interactions between the bare state and the two-particle channels as well as within and between the coupled channels.
However, in this study, most of the energy levels and  phase shifts are in the resonance region, so the dominant interaction is that between the bare state and two-particle channels.
In addition, given the limited data concerning the energy levels from LQCD as well as experimental observables, we find that, in practice, the existing data can be described well without introducing channel-channel interactions. 
For this reason, the $\pi\pi-\pi\pi$, $\pi\pi-\pi\omega$ and $\pi\omega-\pi\omega$ t/u-channel interactions are neglected.

Because of the definite $J^P$ quantum number of the bare $\rho$ meson, it is sufficient to focus on the Hamiltonian in the subspace spanned by $\ket{\alpha=\pi\pi;k^*,J=1,M,\ell=1,S=0}$ and $\ket{\alpha=\omega\pi;k^*,J=1,M,\ell=1,S=1}$.
For convenience, the $JLS$ indices will be suppressed hereafter. 

The free energy part of the Hamiltonian in this subspace, $H_0$, is given by,
\begin{align}
   H_0 &= \sum\limits_{M} m_\rho^\B\ket{\rho_B,\,M}\,\bra{\rho_B,\,M}+\sum_{\alpha,M}\int k^{*\,2}\,d k^* 
\nonumber\\&
\left(E_{\alpha_1}(k^*)+E_{\alpha_2}(k^*)\right)
\ket{\alpha;k^*,\,M}\,\bra{\alpha;k^*,\,M} \, , 
\label{eq:H0}
\end{align}
where $|\,\rho_B,\,M\rangle$ indicates the bare $\rho$ state with $z$-component of spin, $M$, $E_{\alpha_i}(k)=\sqrt{k^2 + m_{\alpha_i}^2}$ with $\alpha_i=\pi$ or $\omega$, 
$m_{\alpha_i}$ is for the mass of the particle in the $\alpha$ channel and  
$m_\rho^\B$ is the mass of the bare single-particle basis state.

The interacting part, $H_I$, is given within the model by,
\begin{align}
   H_I &= \sum_{\alpha,M}\int k^{*\,2}\,d k^*
\left\{\,V_{\alpha}(k^*)
\ket{\rho_B,\,M}\,\bra{\alpha;k^*,\,M}
+\text{h.c.}
\right\} \, , 
\label{eq:HI}
\end{align}
where the interaction term $V_\alpha$ is independent of $M$, as a consequence of the Wigner-Eckart theorem, and given by
\begin{align}
 V_{\pi\pi}(k^*) 
 &= \frac{g_{\rho\pi\pi}}{2\pi\sqrt{3}}
 \frac{k^*}{\sqrt{m_\rho^\B}\,E_{\pi}(k^*)}\,u_{\pi\pi}(k^*) \, , 
 \label{eq:inf Vrhopipi}
 \\
 V_{\omega\pi}(k^*)  
 &= \frac{g_{\omega\rho\pi}}{2\pi\sqrt{6}}\frac{k^* \sqrt{m_\rho^\B}}{\sqrt{E_{\pi}(k^*)\,E_{\omega}(k^*)}}\,u_{\omega\pi}(k^*) \, , 
 \label{eq:inf Vrhoomegapi}
\end{align}
where $u_{\pi\pi}$ and $u_{\omega\pi}$ are the form factors parameterizing the internal structure of hadrons and ensuring the convergence of loop integrals. 
Here, the usual dipole form factors are used~\cite{Matsuyama:2006rp, Leinweber:2001ac},
\begin{align}
    u_{\pi\pi}(k) &= \left(\frac{\Lambda_{\rho\pi\pi}^2}{k^2+\Lambda_{\rho\pi\pi}^2}\right)^2 \, ,
    \label{eq:pipicutoff}
    \\
    u_{\omega\pi}(k) &= \left(\frac{\Lambda_{\omega\rho\pi}^2-\mu_\pi^2}{k^2 + \Lambda_{\omega\rho\pi}^2}\right)^2 \, ,
    \label{eq:omegapicutoff}
\end{align}
where $\mu_\pi =138.5$MeV is the physical mass of the pion.

The scattering T-matrix, defined by $S_{fi} = \delta_{fi} - 2\pi i\, \delta^4(p_f-p_i)\,T_{fi}$, can be obtained from the partial wave Lippmann-Schwinger equation~\cite{Matsuyama:2006rp, Wu:2012md, Wu:2014vma},
\begin{align}
   T_{\alpha\beta}(p,q,E) &= V_{\alpha\beta}(p,q,E)\nonumber\\
   &+\sum_{\gamma}\int k^2\, dk \frac{V_{\alpha\gamma}(p,k,E)\,T_{\gamma\beta}(k,q,E)}{E-E_{\gamma_1}(k)-E_{\gamma_2}(k)+i\varepsilon} \, ,
    \label{eq:LSequation}
\end{align}
where $V_{\alpha\beta}$ comes from bare $\rho$ exchange in the  $s$-channel and is given by
\begin{align}
  V_{\alpha\beta}(p,q,E) &= \frac{V^*_{\alpha}(p)\,V_{\beta}(q)}{E-m^B_\rho} \, .
  \label{eq:potential}
\end{align}
with $V_{\alpha}$ defined in Eqs.(\ref{eq:inf Vrhopipi}) and (\ref{eq:inf Vrhoomegapi}).
In the present case $T_{\pi\pi,\,\pi\pi}(p,q;E)$ can be obtained analytically
\begin{align}
& T_{\pi\pi,\,\pi\pi}(p,q;E) = V^*_{\pi\pi}(p)\,G(E)\,V_{\pi\pi}(q) \, ,
\end{align}
where $G(E)$ is the full propagator of the $\rho$ meson defined by,
\begin{align}
    G(E)^{-1} = E -  m_\rho^\B - \Sigma(E) \, ,
    \label{eq:propagator}
\end{align}
with the self-energy
\begin{align}
\Sigma(E) &= \Sigma_{\pi\pi}(E) + \Sigma_{\omega\pi}(E), \\
    \Sigma_{\pi\pi}(E) &= \int q^2 \,dq\,\frac{\abs{V_{\pi\pi}(q)}^2}{E-2E_{\pi}(q)+i\varepsilon} \, ,  
\label{eq:self pipi}
    \\
    \Sigma_{\omega\pi}(E) &= \int q^2 \,dq\,\frac{\abs{V_{\omega\pi}(q)}^2}{E- E_{\pi}(q) - E_{\omega}(q) + i\varepsilon} \, . 
\label{eq:self omegapi}
\end{align}
The partial-wave phase shift, $\delta(E)$, for the $P$-wave $\pi\pi \to \pi\pi$ elastic scattering is then given by,
\begin{align}
    e^{2i\delta(E)}  &= 1 - i \frac{\pi \bar{p} E}{2} \,\bar{p}\,T_{\pi\pi,\,\pi\pi}(\bar{p} \, ,\bar{p};E),
\\
    \delta(E) &= \arctan\left[\frac{\Im\Sigma_{\pi\pi}(E)}{E-m_\rho^\B -\Re\Sigma(E)}\right] \,\, \left(\mathrm{mod}\,\,\pi\right) \,,
    \label{eq:phaseshift}
\end{align}
where $\bar{p}=\sqrt{E^2/4-m^2_\pi}$ is the on-shell momentum.
The pole position of the $\rho$-resonance is located in the lower half plane of the unphysical Riemann sheet of the $\pi\pi$-channel but the first Riemann sheet of the $\omega\pi$-channel and determined by solving the equation
\begin{align}
    0 = E -  m_\rho^\B - \Sigma(E) \, .
    \label{eq:pole of propagator}
\end{align}
%

\subsection{The Hamiltonian in finite volume}
To obtain the energy levels in finite volume, we need to construct the finite volume Hamiltonian (FVH). Two major problems are encountered. 
Firstly, the correspondence between the Fock spaces spanned by the states with continuous and discrete momentum and secondly, the symmetry is reduced from the $\mathrm{O}(3)$ group to a finite subgroup, $G$, for the finite volume. As a result, $J$ and $M$ are no longer good quantum numbers.
In Refs.\cite{Li:2019qvh, Li:2021mob}, the standard formalism for the rest, moving 
and elongated frames were presented.
Here we give a brief introduction to those aspects relevant to the present work.

To obtain the FVH in terms of the states with discrete momentum, one needs to make the following substitutions in Eqs.(\ref{eq:H0}) and (\ref{eq:HI}). First one sets 
\begin{align}
\ket{\rho_B,\,M} &\to \ket{\rho_B,\,M}_L \, ,
\label{eq:rhoM}
\end{align}
because the bare $\rho$ single-particle state does not change.
However, it is very different for the two-particle state,
\begin{align}
\ket{\alpha;k^*,\,M} &\to \sqrt{\frac{V}{\left(2\pi\right)^3}}\ket{\alpha;e_{\boldsymbol{n}},\,M} \, ,
\label{eq:alphaM}
\end{align}
and
\begin{align}
    \int \dd^3 \boldsymbol{k}^* \to \frac{(2\pi)^3}{V} \sum\limits_{\boldsymbol{n}\in\mathbb{Z}^3} \, ,
\label{eq:int2sum} 
\end{align}
where $V=\eta L^3$ is the volume of the box, with elongation factor $\eta$, and $e_{\boldsymbol{n}}$ denotes a degenerate shell of the non-interacting Hamiltonian in the rest frame, because those states with the same $e_{\boldsymbol{n}}$ share the same 
$|\boldsymbol{k}^*(\boldsymbol{n})|$. 
For example, in the rest frame of a cubic box, 
$\boldsymbol{k}^* = \frac{2\pi}{L}\boldsymbol{n}$ and hence $e_{\boldsymbol{n}}=\boldsymbol{n}^2$. 
However, for the general case, $\boldsymbol{k}^*(\boldsymbol{n})$ and hence $e_{\boldsymbol{n}}$ are not that simple.
A detailed discussion of $e_{\boldsymbol{n}}$ and $\boldsymbol{k}^*(\boldsymbol{n})$ can be found in Ref.~\cite{Li:2021mob} and a summary is given in Appendix~\ref{app:knen}. 
The finite volume basis vector, $\ket{\alpha;e_{\boldsymbol{n}},\,M}$, is given by an expression analogous to Eq.~(\ref{eq:inf JMLS basis})
\begin{align}
   \ket{\alpha; e_{\boldsymbol{n}},M} = &A_\alpha
   \sum\limits_{m\sigma\sigma_1\sigma_2}
   C_{\ell S}(JM;m\sigma)
   \,C_{s_1s_2}(S\sigma;\sigma_1\sigma_2) 
   \notag \\
   &\times 
   \sum\limits_{\boldsymbol{n}\in\{\hat{e}_{\boldsymbol{n}}\}}
   Y_{\ell m}(\hat{\boldsymbol{k}}^*(\boldsymbol{n}))
   \ket{\boldsymbol{k}^*(\boldsymbol{n}),\sigma_1\sigma_2} \, , 
\label{eq:fin JMLS basis}
\end{align}
with $J=1$, $\ell=1$ and $S=0/1$ for $\alpha=\pi\pi / \omega\pi$, respectively. 
Here $\{\hat{e}_{\boldsymbol{n}}\}$ denotes the set of integer vectors with the same $e_{\boldsymbol{n}}$. 

Note that the states defined in Eq.~(\ref{eq:inf JMLS basis}) with different values of $JM\ell S$ are orthogonal, which is not the case in the finite volume since $\mathrm{O}(3)$ symmetry is broken.
Thus, it is necessary to construct an orthogonal basis $\ket{\alpha;e_n,\Gamma,a}$ furnishing an irreducible representation $\Gamma$ of $G$.
Such states take linear combinations of the basis states $\ket{\alpha; e_{\boldsymbol{n}},M}$ with reduction coefficients $C_{\Gamma,\,G}$~\cite{Li:2021mob, Bernard:2008ax},
\begin{align}
   & \ket{\alpha;e_{\boldsymbol{n}},\Gamma,a} := \sqrt{\frac{1}{Z_\Gamma(e_{\boldsymbol{n}})}} \left[C_{\Gamma,\,G}\right]_{M,a}\ket{\alpha;e_{\boldsymbol{n}},M},
    \label{eq:fin Gamma}\\
& \braket{\alpha;e_{\boldsymbol{n}},\Gamma,a}{\alpha';e'_{\boldsymbol{n}},\Gamma',a'} = \delta_{\alpha\alpha'}\,\delta_{e_{\boldsymbol{n}}e'_{\boldsymbol{n}}}\,\delta_{\Gamma\Gamma'}\,\delta_{aa'} \, ,
   \label{eq:normal3}
\end{align}
where $Z_\Gamma(e_{\boldsymbol{n}})$ is the normalization factor.
In general there should be another index denoting the multiplicity of $\Gamma$, but in the present case that additional index is always 1 and hence it will be suppressed.
The reduction coefficients relevant to the work reported here are shown in Appendix ~\ref{app:enstata}.
Similarly,
\begin{align}
\ket{\rho_B,\,\Gamma,\,a}=\left[C_{\Gamma,\,G}\right]_{M,a}\ket{\rho_B,\,M}_L \, ,
\end{align}
satisfying
\begin{align}
   \braket{\rho_B,\,\Gamma,\,a}{\rho_B,\,\Gamma',\,a'}=\delta_{\Gamma\Gamma'}
\delta_{aa'} \, .
\end{align}

With these well-defined orthogonal basis states $\ket{\rho_B,\,\Gamma,\,a}$ and 
$\ket{\alpha;e_{\boldsymbol{n}},\Gamma,a}$, the FVH in the rest frame for a given irreducible representation, $\Gamma$, (note that because of the Wigner-Eckart theorem, the eigenvalue is independent of the ``$a$'' index, which is therefore suppressed) is given by
\begin{align}
H^\text{fin} &= \sum_\Gamma\left( H^\text{fin}_{0,\,\Gamma} + H^\text{fin}_{I,\,\Gamma}\right) \, ,
    \label{eq:fin H}
    \\
    H^\text{fin}_{0,\,\Gamma} &= m_\rho^\B \ket{\rho_\B,\,\Gamma}\bra{\rho_\B,\,\Gamma} + \sum\limits_{\alpha;e_{\boldsymbol{n}}} 
     \ket{\alpha;e_{\boldsymbol{n}},\Gamma}\bra{\alpha;e_{\boldsymbol{n}},\Gamma}
   \nonumber\\&\times
    \left(E_{\alpha_1}(|\boldsymbol{k}^*(\boldsymbol{n})|)
    +E_{\alpha_2}(|\boldsymbol{k}^*(\boldsymbol{n})|)\right) \, ,     
    \label{eq:fin H0}
    \\
    H^\text{fin}_{I,\,\Gamma} &= \sum\limits_{\alpha,e_{\boldsymbol{n}}} V^\text{fin}_{\alpha,\Gamma}(|\boldsymbol{k}^*(\boldsymbol{n})|) \ket{\rho_\B,\,\Gamma}\bra{\alpha;e_{\boldsymbol{n}},\Gamma} + \text{h.c.} \, .
    \label{eq:fin V}
\end{align}
Alternatively, these equations may be expressed in matrix form
\begin{align}
&H^\text{fin}_{0\,\Gamma}+H^\text{fin}_{I\,\Gamma} = 
 \begin{pmatrix}
   m_\rho^\B & v^\mathrm{T} \\
   v  & h_0
 \end{pmatrix}
   \\
&   v^\mathrm{T} = \left( 
   V^\text{fin}_{\pi\pi,\Gamma}(|\boldsymbol{k}^*(\boldsymbol{n}_1)|) , V^\text{fin}_{\pi\pi,\Gamma}(|\boldsymbol{k}^*(\boldsymbol{n}_2)|) ,
   \cdots,
\right.\nonumber\\&\hspace{1.5cm}\cdots,\left.
   V^\text{fin}_{\omega\pi,\Gamma}(|\boldsymbol{k}^*(\boldsymbol{n}'_1)|) , V^\text{fin}_{\omega\pi,\Gamma}(|\boldsymbol{k}^*(\boldsymbol{n}'_2)|) ,\cdots
   \right)
   \\
&  h_0 = \mathrm{diag}\left(
2E_{\pi}(|\boldsymbol{k}^*(\boldsymbol{n}_1)|) , 2E_{\pi}(|\boldsymbol{k}^*(\boldsymbol{n}_2)|),\cdots,
\right.\nonumber\\&\hspace{1.5cm}\left.
E_{\pi}(|\boldsymbol{k}^*(\boldsymbol{n}_1^\prime)|)
+E_{\omega}(|\boldsymbol{k}^*(\boldsymbol{n}_1^\prime)|),\right.\nonumber\\&\hspace{1.5cm}\left.
E_{\pi}(|\boldsymbol{k}^*(\boldsymbol{n}_2^\prime)|)+E_{\omega}(|\boldsymbol{k}^*(\boldsymbol{n}_2^\prime)|),\cdots
\right) \, .
\end{align}
In principle the FVH is countably infinite-dimensional, while in practice it is found that the contribution of high-energy states to the low-lying eigenvalues of interest is negligible. 
Therefore, the matrix can be truncated by excluding the states with momentum higher than a certain value $k^*_\text{cut}$ to obtain a finite-dimensional matrix.

By comparing Eqs.(\ref{eq:inf JMLS basis}),(\ref{eq:HI}),(\ref{eq:fin JMLS basis}),(\ref{eq:fin Gamma}) and (\ref{eq:fin V}), 
$V^\text{fin}_{\alpha,\Gamma}(|\boldsymbol{k}^*(\boldsymbol{n})|)$ is written as,
\begin{align}
    V^\text{fin}_{\alpha,\Gamma}(e_{\boldsymbol{n}}) = \sqrt{\frac{(2\pi)^3}{V}} J_\alpha(e_n) \sqrt{Z_\Gamma(e_n)}V_\alpha(|\boldsymbol{k}^*(\boldsymbol{n})|) \, ,
    \label{eq:matrix element of FVH}
\end{align}
where the Jacobian, $J_\alpha$, just appears for a moving system, with the expression shown in Eq.~(\ref{eq:jacob}).

\subsection{Fitting Formulas}
In the present work, we fit the eigenvalues of the FVH, $E^\text{H}$, to the lattice spectrum, $E^\text{lat}_\text{cm}$, with the usual least-$\chi^2$ strategy. That is, we minimize the 
$\chi^2$ defined as
\begin{align}
    \chi^2 = (E^\text{H} - E^\text{lat}_\text{cm})^\mathrm{T}\mathbb{C}^{-1} (E^\text{H} - E^\text{lat}_\text{cm}) \, ,
\end{align}
where $\mathbb{C}$ denotes the covariance matrix of the lattice spectrum. 
It should be noted that $E^\text{lat}_\text{cm}$ is the spectrum that has been transformed into the rest frame. If a certain energy level, $E^\text{lat}_n$, is extracted from the composite operator with $\boldsymbol{P}\neq0$, it needs to be converted to $E^\text{lat}_{\text{cm},n}$ through
\begin{align}
    E^\text{lat}_{\text{cm},n} = \sqrt{ (E^\text{lat}_{n})^2 - \boldsymbol{P}^2} \, .
\end{align}

In our model, there are five parameters, including the bare mass, $m_\rho^\B$, two coupling constants $g_{\rho\pi\pi}$ and $g_{\omega\rho\pi}$, defined in Eqs.~(\ref{eq:inf Vrhopipi}) 
and (\ref{eq:inf Vrhoomegapi}), and two cut-off parameters $\Lambda_{\rho\pi\pi}$ and 
$\Lambda_{\omega\rho\pi}$, defined in Eqs.~(\ref{eq:pipicutoff}) and 
(\ref{eq:omegapicutoff}), respectively.

\subsection{Formulas for the $m_\pi$ dependence}
To extrapolate the results of the lattice calculations to the physical region, one needs to investigate the $m_\pi$-dependence of the properties of the $\rho$ meson.
This issue has been previously discussed in some studies~\cite{Allton:2005fb, Leinweber:2001ac, Bruns:2004tj, Armour:2005mk,Chen:2012rp,Molina:2020qpw,Niehus:2020gmf} and in this paper it will be studied within the framework of HEFT.
As discussed in Ref.~\cite{Bruns:2004tj}, in the framework of Chiral Perturbation Theory, the mass of the $\rho$ is a function of $m_\pi$ of the form
\begin{align}
    m_\rho^\p = c^\prime_0 + c^\prime_1 \, m_\pi^2 + c^\prime_2 \, m_\pi^3 + c^\prime_3 \, m_\pi^4 \ln\left(\frac{m_\pi^2}{m_\rho^2}\right) + \mathcal{O}(m_\pi^4) \, ,
\label{eq:mpi dependence of mrho perturbation theory}
\end{align}
where $m_\rho^\p$ is related to the pole mass of the $\rho$, corresponding to the real part of the pole of the T-matrix in the complex plane.
In general, $m_\rho^\p$ is different from the usual Breit-Winger mass, $m_\rho^{\text{BW}}$,  which is defined as the real energy at which the phase shift is 90 degrees.
For the present case, however, the difference is negligible. 

As discussed in Refs.~\cite{Leinweber:2001ac,Bruns:2004tj}, the quark mass insertion at tree level only contributes to the $m_\pi^2$ term in Eq.~(\ref{eq:mpi dependence of mrho perturbation theory})  up to $\mathcal{O}(m_\pi^4)$, with the other two terms arising from pion-loop self energies.
The $m_\pi^3$ term comes from the $\omega\pi$ loop, involving a vector-vector-pseudoscalar ($VVP$) vertex, while both $\pi\pi$ and $\omega\pi$ loops contribute to the $\log$ term.  Within 
the present framework, the pole of the T-matrix is determined by Eq.~(\ref{eq:pole of propagator}), which tells us that
\begin{align}
    m_\rho^\p = m_\rho^\B + \Re\Sigma(m_\rho^\p - i\Gamma/2) \, .
    \label{eq:mrho approxiamte condition HEFT}
\end{align}
%

where $m_\rho^\p-i\Gamma/2$ is the complex pole position solved from Eq.~(\ref{eq:pole of propagator}). 
Comparing Eq.~(\ref{eq:mpi dependence of mrho perturbation theory}) and Eq.~(\ref{eq:mrho approxiamte condition HEFT}), the bare mass $m_\rho^\B$ is a quadratic function of $m_\pi$ at the leading order, 
\begin{align}
    m_\rho^\B(m_\pi) = c_0 + c_1 \, m_\pi^2 \, .
    \label{eq:mpi depdence of mrhoB}
\end{align}
In our analysis, this equation will be used to study the extracted $m_\rho^\B$ as function of $m_\pi$.

It is important to note that Eq.~(\ref{eq:mpi depdence of mrhoB}) is derived from the continuum field theory. 
In principle, to extrapolate the results obtained from lattice, the residual lattice artifacts should be estimated and removed. 
Since all actions are $O(a)$-improved, the effect of the finite lattice spacing can be estimated from $O(a^2)$ as
\begin{align}
    m_\rho^\B(m_\pi;a) = c_0 + c_1 m_\pi^2 + \xi a^2 \, ,
    \label{eq:mpi depdence of mrhoB with lat spacing}
\end{align}
where $\xi$ characterizes the rate at which it approaches the continuum limit and may vary from collaboration to collaboration as different fermion actions are used.


\subsection{Model (in)dependence in HEFT}

Understanding the model-dependent and model-independent aspects of HEFT is important.
As HEFT incorporates the L\"uscher formalism
\cite{Wu:2014vma, Hall:2013qba}, there are aspects of the calculation that share the same level of model independence as the L\"uscher formalism itself.

\subsubsection{Model independence}

In particular, the L\"uscher formalism provides a rigorous relationship between the finite volume energy spectrum and the scattering phase shifts and inelasticities of infinite-volume experiment.  
In HEFT, this relationship is mediated by a Hamiltonian. 
When the parameters of the Hamiltonian are sufficient to provide a high-quality
description of lattice QCD results, then the associated scattering amplitudes are of high quality.
The key is to have a sufficient number of tunable parameters to accurately describe the lattice QCD results.

In the baryon sector, high-quality lattice QCD results are scarce and HEFT is often fit to experimental data first.
The HEFT formalism then describes the finite-volume dependence of the hadronic spectrum, indicating where future lattice QCD results will reside.

Fortunately, in the $\rho$-meson channel relevant to this analysis, several lattice QCD groups have resolved the finite-volume spectrum, taking care to assess the subtle shifts in the spectrum associated with avoided level crossings in the finite volume.  
This information is central to the L\"uscher formalism and as such, is central to the HEFT analysis presented here.
We will show excellent fits to the lattice
QCD results such that HEFT provides rigorous predictions of the scattering observables with model independence at the level of the L\"uscher formalism.

Of course, this model independence is restricted to the case of matched quark masses in finite-volume and infinite-volume.  
The L\"uscher formalism provides no avenue for changing the quark mass.
In other words, to make contact with experiment, the quark masses used in the lattice QCD simulations must be physical.

On the other hand, $\chi$PT is renowned for describing the quark mass dependence of hadron properties in a model-independent manner, provided one employs the truncated expansion in the power-counting regime, where higher-order terms not considered in the expansion are small by definition.
Given that finite-volume HEFT reproduces
finite-volume $\chi$PT in the perturbative limit by construction~\cite{Hall:2013qba, Abell:2021awi}, it is reasonable to explore the extent to which this model independence persists in the full nonperturbative calculation of HEFT.

This was explored in Ref.~\cite{Abell:2021awi}.
In the one channel case where a single particle basis state (e.g. a quark-model like $\Delta$) couples to one two-particle channel (e.g. $\pi N$), the independence of the results on the form of regularisation is
reminiscent of that realised in $\chi$PT.
Any change in the regulator is absorbed by the low-energy coefficients such that the renormalised coefficients are physical, independent of the renormalisation scheme. 

However, in the more complicated two-channel case with a $\pi \Delta$
channel added, the same was not observed.  The form of the Hamiltonian
becomes constrained, describing experimental data accurately
for only a limited range of parameters.  The Hamiltonian becomes a
model in this case, with regulator-function scales and shapes governed
by the experimental data.  The principles of chiral {\em perturbation}
theory no longer apply in this nonperturbative calculation.  However, for fit
parameters that describe the data well, the model independence of the
L\"uscher formalism remains intact.  The Hamiltonian is only mediary.

\subsubsection{Quark mass variation}

The consideration of variation of the quark masses away from the physical point provides further constraints on the Hamiltonian.
In particular, lattice QCD results away from the physical point provide new constraints on the form of the Hamiltonian.
In the two-channel case, the Hamiltonian
becomes tightly constrained when considering experimental scattering data
and lattice QCD results together.  

With the Hamiltonian determined by one set of lattice results, one can then make predictions of the finite-volume spectrum considered by other lattice groups at different volumes and different quark masses.
For the cases considered in the baryon spectrum the predictions of HEFT are in agreement with lattice QCD spectrum predictions.  
For example, in the $\Delta$-channel HEFT successfully predicts the finite-volume spectrum of the CLS consortium~\cite{Abell:2021awi, Morningstar:2021ewk}.
In the $N(\frac{1}{2}^+)$ channel,
HEFT reproduces the lattice QCD results from Lang {\it et al.}
\cite{Wu:2017qve, Lang:2016hnn}.  
In the $N(\frac{1}{2}^-)$ channel, HEFT
successfully predicts spectra from the CLS consortium
\cite{Abell:2023nex,Bulava:2022vpq}, the HSC
\cite{Abell:2023nex,Edwards:2011jj,Edwards:2012fx} and Lang \&
Verducci \cite{Abell:2023nex, Lang:2012db}. 
Thus one concludes that the systematic errors of the HEFT approach to quark-mass variation are small on the scale of contemporary lattice QCD uncertainties.
As the Hamiltonian is constrained by model-independent scattering data and lattice QCD results, we expect this success to be realised in the $\rho$-meson channel.

Variation in the quark mass is conducted in the same spirit as for $\chi$PT.
The couplings are held constant and the hadron masses participating in the theory take values as determined in lattice QCD.
The single-particle bare basis state acquires a quark mass dependence and this is done in the usual fashion by drawing on terms analytic in the quark mass.  
In most cases, lattice QCD results are
only able to constrain a term linear in $m_\pi^2$, but on occasion, the data can demand a small $m_\pi^4$ contribution.  

In the present analysis, we will see that the accuracy of contemporary lattice QCD results for the $\rho$-meson spectrum is sufficient to consider only a term linear in $m_\pi^2$.
Even then we show that there are incompatibilities between the lattice QCD results.
Contrary to the baryon sector, we show how a Hamiltonian constrained by the results of one group is incompatible with the results of other groups.
Referring back to the model-independence of the L\"uscher relation embedded within HEFT, the spectra of one group leads to scattering observables that are incompatible with the predictions of other
groups.

The model independence associated with the movement of quark masses away from the physical point is largely governed by the distance one chooses to move from the physical quark-mass point.
The HEFT approach is systematically improvable, reliant on high-quality lattice QCD results to constrain the higher-order terms that one can introduce.
In addition to the aforementioned analytic $m_\pi^4$ term, one could also include higher-order interaction terms from the chiral Lagrangian.
However, this increased level of precision is not yet demanded by contemporary lattice QCD results.

\subsubsection{Model dependence}

Now that the Hamiltonian has become a tightly constrained model, the eigenvectors describing the manner in which the non-interacting basis states come together to compose the eigenstates of the spectrum are model dependent. 
At the same time, there is little freedom in the model parameters of the Hamiltonian such that the predictions of the Hamiltonian are well defined.

With regard to the bare mass, there is interplay between the multi-particle channels included in the calculation, the
regularisation scales considered and the associated bare mass.
For example, as will be seen, upon introducing the additional $\omega \pi$ channel, one observes an increase in the bare mass as some of this contribution is
carried by the $\omega \pi$ contribution.  
Thus the bare mass is defined only when the number of channels and their regularisation
scales are fixed within the model.
With the channels selected and preferred regularisation scales set, the bare mass becomes well-defined within the model.  

While the bare mass is model dependent, we will show that the slope of the bare mass as a function of quark mass is insensitive to the number of two-particle channels considered.  
As such, this observation may be of assistance to those developing quark models.

Returning to the eigenvectors of the Hamiltonian, we emphasise that the parameters of the Hamiltonian model are well constrained, such that the predictions of the model are well defined.

The information contained in the Hamiltonian eigenvectors describing the basis-state composition of finite-volume energy eigenstates is analogous to the information contained within the  eigenvectors of lattice QCD correlation matrices describing the linear combination of interpolating fields isolating energy eigenstates on the lattice.
These too are model dependent, governed by the nature of the interpolating fields used to construct the correlation matrix.  

What is remarkable is that with a suitable renormalisation scheme on the lattice (e.g. interpolators are normalised to set diagonal
correlators equal to 1 at one slice after the source), the composition of the states drawn from the lattice correlation matrix is very similar to the description provided by HEFT~\cite{Wu:2017qve,Abell:2023nex}. 
While both eigenvector sets are model
dependent, their similarity does indeed provide some relevant insight into hadron structure.
And because regularisation in the Hamiltonian is tightly constrained, one can begin to separate out the contributions of bare versus two-particle channels, something that is impossible in $\chi$PT.

\subsubsection{Summary}

In summary, there is a direct model-independent link between the finite-volume spectrum calculated at physical quark masses and the scattering observables of experiment.
This model independence is founded on the L\"uscher formalism embedded with HEFT.
Similarly, variation of the quark masses away from the physical quark mass has systematic uncertainties that are small relative to contemporary lattice QCD spectral uncertainties.
Finally, the Hamiltonian eigenvectors describing the basis-state composition of finite-volume energy eigenstates are model dependent.
They are analogous to the interpolator dependent eigenvectors of lattice QCD correlation matrices describing the linear combination of interpolating fields isolating energy eigenstates on the lattice.
The similarity displayed by these two different sets of eigenvectors suggests that they do indeed provide insight into hadron structure.

\section{Numerical Results and Discussion}
\label{sec:discussion}

\subsection{The LQCD Data}
The finite volume spectra for the $I=\ell=1$ $\pi\pi$ sector with dynamical fermions at various pion masses have been provided by several LQCD collaborations over the past decade , including PACS-CS (2011, $N_f=2+1$)~\cite{CS:2011vqf}, HSC (2013, $N_f=2+1$)~\cite{Dudek:2012xn}, HSC (2015, $N_f=2+1$)~\cite{Wilson:2015dqa}, Guo \textit{et al}. (2016, $N_f=2$)~\cite{Guo:2016zos}, MILC(2016, $N_f=2+1$)~\cite{Fu:2016itp}, C. Alexandrou \textit{et al}. (2017, $N_f=2+1$)~\cite{Alexandrou:2017mpi}, J. Bulava \textit{et al}. (2019, $N_f=2+1$)~\cite{Andersen:2018mau}, and ETMC (2020, $N_f=2+1(+1)$)~\cite{ExtendedTwistedMass:2019omo}. 
Further details and energy levels are shown in Table~\ref{tab:detail of spectrum} and the panels in Fig.~\ref{fig: spectra example}, as well as all figures in Appendix~\ref{app:fiting levels}, where HEFT is fit to the various lattice data sets.


\begin{table*}
\caption{Details of the spectra by different collaborations. Columns, from left to right, show collaboration and year of spectra, pion mass ($m_\pi$), number of flavor ($N_f$), lattice size ($L$) and spacing ($a$) in fm, employed gauge and fermion actions, number of energy levels ($N_{\text{lvl}}$), and energy level extraction method. Besides, the spectra provided by Guo \textit{et al}. are extracted in a elongated box with factor$\eta=1, \frac{7}{6}, \frac{4}{3}$ for $m_\pi=226$ MeV, and $\eta=1, 1.25, 2$ for $m_\pi=315$ MeV.}
\begin{tabular}{cccc@{\hspace{12pt}}c@{\hspace{12pt}}ccc}
    \toprule
  COLLAB.(Year) & $m_\pi$(MeV) &  $N_f$ & $L$(fm) & $a$(fm) & Action & $N_{\text{lvl}}$ & method  \\
  \midrule
  \multirow{3}{*}{J.Bulava(2018)} & 200 & \multirow{3}{*}{2+1} & 4.1 & 0.06 & \multirow{3}{*}{\makecell{improved L\"{u}scher-Weisz gauge\\improved Wilson fermion}} & 17 & \multirow{3}{*}{GEVP} \\ 
                                  & 220 & & 4.1 & 0.09 & &  21 & \\
                                  & 280 & & 3.1 & 0.06 & &  15 & \\
  \midrule
   \multirow{3}{*}{HSC(2013)} & \multirow{3}{*}{391} & \multirow{5}{*}{2+1} &  1.9 & 0.12 & \multirow{5}{*}{\makecell{Symanzik-improved gauge\\anisotropic Clover fermion}} &  7 & \multirow{4}{*}{GEVP} \\
                              & & & 2.4 & 0.12 &  & 10 & \\
                              & & & 2.9 &0.12 & &  14 & \\
  HSC(2015) & 236 &  & 3.8 & 0.12 &  &  23 &  \\
  HSC(2023) & 330 & & 2.8 & 0.12 &  &  17 & \\
  \midrule 
  \multirow{6}{*}{MILC(2016)}  &  176 & \multirow{6}{*}{2+1} & 5.4 & 0.09 & \multirow{6}{*}{\makecell{improved L\"{u}scher-Weiss gauge\\staggered fermion}} & 9 & \multirow{6}{*}{GEVP} \\
                          &  247 & & 3.4 & 0.09 & &  9 & \\
                          &  248 & & 3.4 & 0.09 & &  9 & \\
                          &  301 & & 2.7& 0.09 & &  9 & \\
                          &  346 & & 2.4 & 0.09 & &  7 & \\
                          &  276 & & 3.7& 0.12 & &  9 & \\ 
  \midrule
  \multirow{5}{*}{ETMC(2020)} & 322 & \multirow{5}{*}{2+1+1} & 2.8 & 0.09 & \multirow{5}{*}{\makecell{Iwasaki gauge\\twisted-mass Wilson fermion}} & 18 & \multirow{5}{*}{GEVP}\\
                        &  386 & & 2.1 & 0.09 & &  16 & \\
                        &  262 & & 2.6 & 0.08 & &  13 & \\
                        &  302 & & 3.9 & 0.08 & &  23 & \\
                        &  376 & & 2.6 & 0.08 & &  14 & \\
  \midrule
  C.Alexandru(2017) & 316 & 2+1 & 3.6 & 0.11 & \makecell{Symanzik-improved gauge\\clover Wilson fermion}  & 15 & GEVP\\
  \midrule
  PACS-CS(2011) & 411 & 2+1 & 2.9 & 0.09 & \makecell{Iwasaki gauge\\improved Wilson fermion} & 6 & Exp Fit \\
  \midrule
  \multirow{2}{*}{Guo(2016)} & 226 & \multirow{2}{*}{2} & \multirow{2}{*} {2.9} & \multirow{2}{*} {0.12} & \multirow{2}{*}{\makecell{L\"{u}scher-Weiss gauge \\ nHYP-smeared Clover fermion}} & 8 & \multirow{2}{*}{GEVP} \\
                        &  315 & & & & & 20 & \\
  \bottomrule
\end{tabular}
\label{tab:detail of spectrum}
\end{table*}

\subsection{Three Fitting Schemes}
\label{sec:fitting method}
In this work, our aim is to study the properties of the $\rho$ meson by investigating the pion mass dependence of various relevant variables.
In the HEFT framework, we have five free parameters which may be used to fit the lattice data with different pion masses. 
However, for each specific pion mass, there are only a few energy levels.
In addition, the $\omega\pi$ contribution is considerably weaker than that of the $\pi\pi$ channel,  since the threshold of $\omega\pi$ is higher than the $\rho$ mass.
Consequently, we first adopt scheme A, wherein the interaction $V_{\omega\pi}$ is turned off, i.e., $g_{\omega\rho\pi}\equiv 0$, while $m_\rho^\B$, $g_{\rho\pi\pi}$, and $\Lambda_{\rho\pi\pi}$ are treated as free fitting parameters.
The finite volume spectra provided by various collaborations involving different $\pi$ masses are each fit independently. 
Using scheme A, it is found that both $g_{\rho\pi\pi}$ and $\Lambda_{\omega\rho\pi}$ show a very weak dependence on $m_\pi$, while $m_\rho^\B$ is strongly dependent on $m_\pi$, as anticipated earlier.

Building upon the results found using scheme A, in scheme B both $g_{\rho\pi\pi}$ 
and $\Lambda_{\rho\pi\pi}$ are fixed to be independent of $m_\pi$, in accord with standard practice in chiral effective field theory. $m_\rho^\B$ is allowed to vary. 
As a result, this approach effectively combines spectra from various pion masses provided by different LQCD groups together in a unified analysis to constrain the variation of the bare mass, $m_\rho^\B$, with $m_\pi$. 

Finally, the contribution of the $\omega\pi$ channel is examined in scheme C.
In this case $V_{\omega\pi}$ is switched on, however, the two coupling constants, $g_{\rho\pi\pi}$ and $g_{\omega\rho\pi}$, as well as two cutoffs, $\Lambda_{\rho\pi\pi}$ and $\Lambda_{\omega\rho\pi}$, are taken to be independent of $m_\pi$, 
while only $m_\rho^\B$ is permitted to vary in the fitting.
These three schemes are summarized in Table~\ref{tab:fitting strategies}.

\begin{table}[ht]
\caption{Outline of the fitting schemes.}
\begin{tabular}{p{1cm}<{\centering}|p{2cm}<{\centering}|p{2cm}<{\centering}|p{2cm}<{\centering}}
\hline
 & $m_{\rho}^\B$ & $g_{\rho\pi\pi},\Lambda_{\rho\pi\pi}$ & $g_{\omega\rho\pi},\Lambda_{\omega\rho\pi}$ \\ \hline
A & free  & free  &  off    \\ 
B & free  & fixed &  off    \\ 
C & free  & fixed & fixed \\ \hline
\end{tabular}
\label{tab:fitting strategies}
\end{table}


\subsection{Fitting Results}
\label{sec:fitresult}

\subsubsection{Results for Scheme A}
\begin{figure}
    \centering
    \includegraphics[width=.5\textwidth]{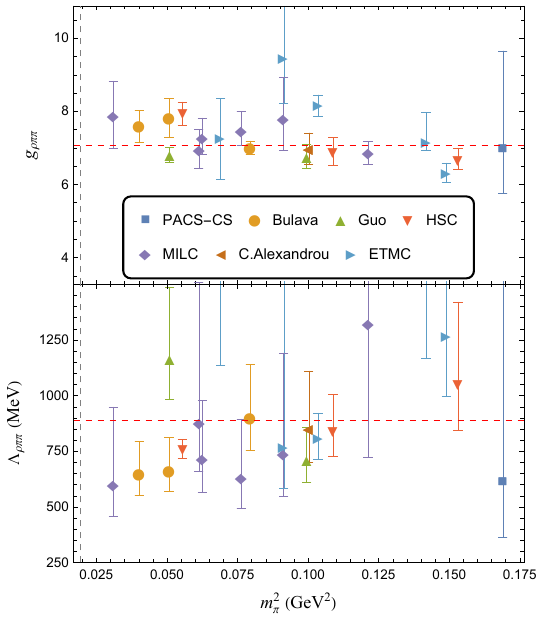}
    \caption{
    $m_\pi$-dependence of $g_{\rho\pi\pi}$ and $\Lambda_{\rho\pi\pi}$ in scheme A. The dashed gray vertical line indicates the physical $m_\pi$. The dashed red horizontal lines indicate the value that will be fixed for $N_f =2+1$ in scheme B.
    }
\label{Fig:FreeFitResult}
\end{figure}

Recall that in scheme A, the  $\rho-\omega\pi$ interaction is turned off and the lattice spectra are fit using three free parameters $m_\rho^\B$, $g_{\rho\pi\pi}$, 
and $\Lambda_{\rho\pi\pi}$. 
The fitted spectra as a function of spatial extent, $L$, are shown by the blue curves in Fig.~\ref{fig: spectra example} for $m_\pi=200$ MeV as an illustration. All other fits are presented in Appendix~\ref{app:fiting levels}.
The corresponding fitted parameter values are outlined in the columns dedicated to scheme A in Table~\ref{tab:Fit Results Table}.
In addition, the pion mass dependence of the three parameters, $g_{\rho\pi\pi}$, $\Lambda_{\rho\pi\pi}$, and $m_\rho^\B$ are shown in Fig.~\ref{Fig:FreeFitResult} and in the top panel of Fig.~\ref{Fig:Result-m}, respectively.

\begin{figure*}[htbp]
    \centering
    \includegraphics[width=0.8\textwidth]{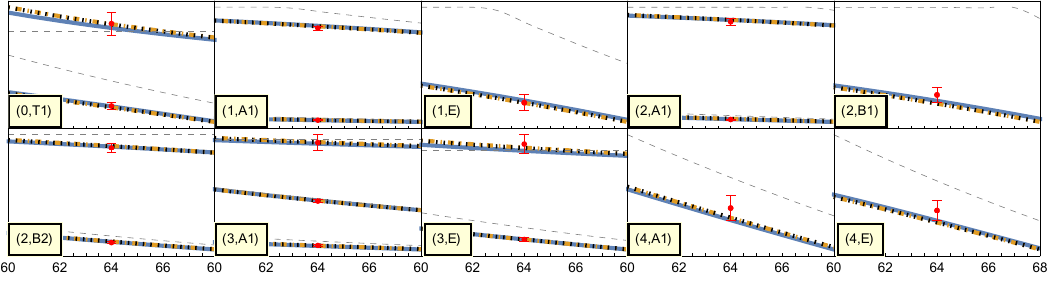}
    \caption{
  Spectra with $m_\pi=200$ MeV provided by Bulava \textit{et al}.~\cite{Andersen:2018mau} along with that calculated by HEFT using the fitting results for schemes A, B and C. 
  The $x$-axis represents the spatial extent $L$ in units of lattice spacing $a$, while the y-axis indicates the energy level. Tick marks on the y-axis are omitted for clarity. Text within the yellow box $(\boldsymbol{n}^2,\Gamma)$ signifies spectrum extraction using operators in representation $\Gamma$ and with total momentum $\boldsymbol{P}^2 = (\frac{2\pi}{L})^2\boldsymbol{n}^2$. 
  Red points indicate the lattice spectrum provided by collaborations.   
  Blue, orange dot-dashed and black dotted lines indicate the spectrum as the function of $L$ calculated by HEFT using the fitting results of schemes A, B and C, respectively. The dashed gray lines indicate the non-interacting energy levels $2E_\pi(\boldsymbol{k^*})$ and $m_\rho^\B$ (with $m_\rho^\B$ taken from Scheme A's fitting result for illustration). The turning points in the non-interacting energy levels are associated with energy crossings where the non-interacting two-particle energy becomes lower than the bare $\rho$-meson mass as $L$ increases.}
  \label{fig: spectra example}
\end{figure*}
%

\begin{table*}
\caption{Fitting results of schemes A, B, and C provided by the MINUIT2 program~\cite{James:1975dr}. 
In the columns for scheme A, we show asymmetric uncertainties, since in this scheme the upper and lower uncertainties are quite different. 
The question mark means that some upper uncertainties are not provided by MINUIT2 since even for a very large $\Lambda_{\rho\pi\pi}$ the energy levels can still be fit well. 
For scheme B, $g_{\rho\pi\pi}$, and $\Lambda_{\rho\pi\pi}$ are fixed at $7.07\,(6.85)$ 
and $890\,(950)$MeV for $N_f=2+1\,(2)$, respectively. 
For scheme C, $g_{\omega\rho\pi}$ 
and $\Lambda_{\omega\rho\pi}$ are fixed at $18$/GeV and $900$ MeV, while $g_{\rho\pi\pi}$ and $\Lambda_{\rho\pi\pi}$ are fixed at $7.07\,(6.85)$ and $900\,(980)$ MeV for $N_f=2+1\,(2)$, respectively. 
$\hat{\chi}^2$ represents the reduced chi-square, i.e, $\chi^2/\text{dof}$ with $\text{dof}=N_{\text{lvl}}-3$ for scheme A and $N_{\text{lvl}}-1$ for schemes B and C.
    }
    \begin{tabular}{ccc@{\hspace{5pt}}c@{\hspace{5pt}}c@{\hspace{5pt}}c@{\hspace{8pt}}c@{\hspace{5pt}}c@{\hspace{8pt}}c@{\hspace{5pt}}c}
    \toprule
  \multirow{2}{*}{COLLAB.(Year)} &  \multirow{2}{*}{$m_\pi$(MeV)} & 
  \multicolumn{4}{c}{Scheme A} & \multicolumn{2}{c}{Scheme B} & \multicolumn{2}{c}{Scheme C} \\
  \cmidrule(lr){3-6}  \cmidrule(lr){7-8}  \cmidrule(lr){9-10} 
  & & $m_\rho^\B$(MeV) & $g_{\rho\pi\pi}$ & $\Lambda_{\rho\pi\pi}$(MeV) & $\hat{\chi}^2$ & $m_\rho^\B$(MeV) & $\hat{\chi}^2$ & $m_\rho^\B$(MeV) & $\hat{\chi}^2$\\

  \midrule
  \multirow{3}{*}{J.Bulava(2018)} 
     & 200 & $ 787.3\aerror{23.6}{14.8} $ & $7.58\aerror{0.45}{0.43}$ & $645.6\aerror{149.8}{93.0}$ & 0.44 & 819.6(4.1) & 0.53 & 867.7(4.5) & 0.53 \\
     & 220 & $795.0\aerror{18.6}{11.5}$ & $7.80\aerror{0.55}{0.51}$ & $662.0\aerror{149.4}{93.0}$ & 0.41 & 818.4(3.6) & 0.49 & 866.3(4.0) & 0.50\\
     & 280 & $825.1\aerror{29.4}{14.7}$ & $7.00\aerror{0.19}{0.17}$ & $895.7\aerror{244.1}{141.7}$ & 1.07 & 826.2(2.4) & 0.93 & 870.8(2.6) & 0.97 \\
     \midrule
     HSC(2013) & 391 & $909.0\aerror{36.4}{15.5}$ & $6.66\aerror{0.33}{0.24}$ & $1050.3\aerror{370.1}{207.8}$ & 0.98 & 898.9(1.2) & 0.97 & 936.2(1.3) & 1.21\\
     HSC(2015) & 236 & $829.1\aerror{29.4}{14.7}$ & $7.94\aerror{0.32}{0.31}$ & $756.4\aerror{46.4}{39.3}$ & 1.00 & 840.3(1.0) & 1.38 & 888.6(1.1) & 1.62 \\ 
     HSC(2023) & 330 & $856.1\aerror{13.1}{8.2}$ & $6.89\aerror{0.41}{0.36}$ & $838.6\aerror{165.9}{112.3}$ & $0.71$ & $862.6(2.2)$ &  $0.78$ & $904.1(2.4)$ & $0.86$ \\
     \midrule
     \multirow{6}{*}{MILC(2016)} 
     & 176 & $806.4\aerror{36.3}{8.5}$ & $7.88\aerror{0.96}{0.88}$ & $596.2\aerror{352.5}{139.1}$ & 0.63 & 831.7(8.2) & 0.61 & 882.9(9.1) & 1.30\\
     & 247 & $878.2\aerror{84.1}{21.2}$ & $6.93\aerror{0.58}{0.49}$ & $876.0\aerror{633.0}{213.5}$ & 0.73& 881.3(3.9) & 0.57 & 928.7(4.3) & 0.53\\
     & 248 & $855.8\aerror{28.6}{12.4}$ & $7.27\aerror{0.55}{0.44}$ & $711.7\aerror{266.1}{144.3}$ & 0.37 & 874.3(4.1) & 0.35 & 921.5(4.5) & 0.32 \\
     & 275 & $861.5\aerror{29.1}{12.5}$ & $7.45\aerror{0.54}{0.39}$ & $630.2\aerror{264.7}{136.7}$ & 0.36 & 886.0(5.5) & 0.43 & 932.4(6.0) & 0.41\\
     & 301 & $916.6\aerror{47.2}{13.8}$ & $7.79\aerror{1.15}{0.86}$ & $735.8\aerror{455.7}{186.0}$ & 0.26 & 928.5(5.8) & 0.29 & 972.4(6.4) & 0.28\\
     & 346 & $999.8\aerror{?}{51.9}$ & $6.86\aerror{0.31}{0.30}$ & $1321.0\aerror{?}{595.8}$  & 0.26 & 958.9(6.4) & 0.23 & 998.6(6.9) & 0.28\\
     \midrule
     \multirow{5}{*}{ETMC(2020)} & 322 & $927.4\aerror{14.4}{10.6}$ & $8.15\aerror{0.30}{0.28}$ & $807.2\aerror{115.0}{91.6}$ & 1.20 & 924.4(2.6) & 2.16 & 967.6(2.8) & 2.06\\
     & 386 & $999.6\aerror{59.5}{21.8}$  & $6.30\aerror{0.28}{0.23}$ & $1267.3\aerror{557.2}{271.1}$ & 1.32 & 974.1(2.2) & 1.54 & 1012.1(2.4) & 2.00\\
     & 262 & $1156.2\aerror{68.7}{208.0}$ & $7.26\aerror{1.10}{1.13}$ & $2421.0\aerror{?}{1283.4}$ & 0.58 & 920.2(8.6)& 0.67 & 967.3(9.5) & 0.72\\
     & 302 & $928.7\aerror{157.1}{23.7}$ & $9.46\aerror{1.42}{1.25}$ & $765.6\aerror{862.6}{182.3}$ & 0.56 & 919.1(5.8)& 1.00 & 964.8(6.3) & 0.99\\
     & 376 & $1034.6\aerror{170.3}{51.8}$ & $7.16\aerror{0.81}{0.23}$ & $1571.4\aerror{?}{403.8}$ & 0.98 & 952.6(1.8)& 1.57 & 993.4(1.9) & 1.52\\
     \midrule
     C.Alexrandru(2017) & 316 & $840.6\aerror{25.8}{12.7}$ & $6.96\aerror{0.45}{0.40}$ & $847.4\aerror{263.9}{147.1}$ &  0.10 & 846.6(2.3) & 0.10 &  890.1(2.5) & 0.10\\
     \midrule
      PACS-CS(2011) & 411 & $913.0\aerror{185.0}{10.2}$ & $6.97\aerror{2.70}{1.21}$ & $609.0\aerror{?}{246.8}$ & 0.83 & 934.3(5.2) & 0.82 & 973.3(5.6) & 0.88\\ 
      \midrule
      \multirow{2}{*}{Guo(2016)}
      & 226 & $834.4\aerror{43.7}{21.3}$ & $6.79\aerror{0.23}{0.19}$ & $1163.1\aerror{323.4}{179.6}$ & 1.69 & 854.3(1.4) & 1.88 & 806.4(1.3) & 1.78 \\
      & 315 & $824.1\aerror{11.6}{7.0}$ & $6.75\aerror{0.35}{0.30}$ & $709.5\aerror{148.2}{97.1}$ & 0.20 & 892.4(1.4) & 1.50 & 848.0(1.3)  & 1.84\\
     \bottomrule
    \end{tabular}

\label{tab:Fit Results Table}
\end{table*}

From Fig.~\ref{Fig:FreeFitResult}, it is found that the cut-off parameter $\Lambda_{\rho\pi\pi}$ and the dimensionless coupling constant, $g_{\rho\pi\pi}$, both show a weak dependence on $m_\pi$.
In contrast, it is worth noting that for each $m_\pi$, $\Lambda_{\rho\pi\pi}$ exhibits a large uncertainty. 
Indeed, in some cases the upper uncertainty does not display in MINUIT2, which means that even for very large values of the cut-off, we can still find a reasonable fit for the energy levels. 
This observation suggests that the eigenvalue of the FVH around the region of the $\rho$ mass is insensitive to $\Lambda_{\rho\pi\pi}$. 
This is consistent with the findings reported in Ref.~\cite{Abell:2021awi}, where the $\Lambda$-dependence of the eigenvalue and eigenvector are investigated in detail.
A more common~\cite{Dudek:2012xn} dressed coupling constant, 
$g^{\text{BW}}_{\rho\pi\pi}$, is defined by
\begin{align}
    g^{\text{BW}}_{\rho\pi\pi} \equiv \sqrt{\frac{6\pi (m_\rho^{\text{BW}})^2 }{\bar{p}} \,\Gamma^{\text{BW}}_{\rho\to\pi\pi}} \, ,
\end{align}
where $\bar{p}$ is defined by $2\sqrt{\bar{p}^2+m_\pi^2}=m_\rho^{\text{BW}}$ and $\Gamma_{\rho\to\pi\pi}^{\text{BW}}$ is the partial width of the resonance in the Breit Wigner parameterizations.
In our model, $g^{\text{BW}}_{\rho\pi\pi}$ depends on both the value of $g_{\rho\pi\pi}$ and the form factor $u_{\pi\pi}$, which is determined by $\Lambda_{\rho\pi\pi}$.
As a result, the slight $m_\pi$-dependence seen in $g_{\rho\pi\pi}$ and $\Lambda_{\rho\pi\pi}$ clearly implies that the value of $g^{\text{BW}}_{\rho\pi\pi}$, approximated to be around $6.0$, remains independent of $m_\pi$.
This conclusion is consistent with the discussions presented in prior works, such as Refs.~\cite{Feng:2010es, Fu:2016itp, Alexandrou:2017mpi}.

In contrast, the $m_\pi$-dependence of $m_\rho^\B$ suffers a large fluctuation, as illustrated in the top section of Fig.~(\ref{Fig:Result-m}). 
This behavior stems from the limited number of energy levels available for each pion mass, coupled with the considerable uncertainties associated with $g_{\rho\pi\pi}$ and $\Lambda_{\rho\pi\pi}$. 
In order to reduce the uncertainty associated with the bare mass, one needs to find an appropriate method to combine the analysis of the various energy levels. 
Inspired by the weak dependence on $m_\pi$, as well as the sizable uncertainties of $g_{\rho\pi\pi}$ and $\Lambda_{\rho\pi\pi}$, we choose these two parameters to be constant for all pion masses and re-fit each spectrum using only one single parameter, $m_\rho^\B$. 
%
%

\subsubsection{Results for Scheme B}

\begin{figure}[tbp]
    \centering
    \includegraphics[width=.45\textwidth]{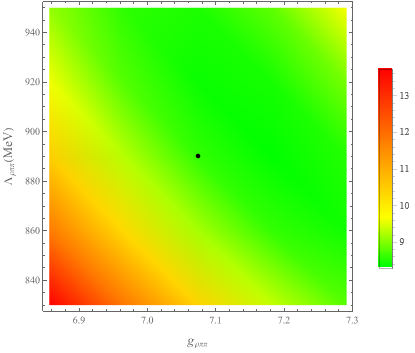}
    \caption{Distribution of $\sum\limits_{m_\pi\in S}\hat{\chi}^2(m_\pi)$ for scheme B using fixed values of $(g_{\rho\pi\pi}$ and $\Lambda_{\rho\pi\pi})$, where 
$S=\{301$,$302$,$315$,$316$,$322\}$. Using the black point marked in the green region ensures 
that $\hat{\chi}^2(m_\pi)\lesssim 2$ is satisfied for each value of $m_\pi\in S$. 
    }
\label{fig:fixgL}
\end{figure}

The first task is to determine the constant values of $g_{\rho\pi\pi}$ and $\Lambda_{\rho\pi\pi}$ which will be used in the analysis of the data for all values of  $m_\pi$. 
To do that, we investigate
$m_\pi=236$, $280$, $322$, $376$, $386$, $391$ MeV cases, which have the highest values of the chi-squared per degree of freedom, $\hat{\chi}^2(=\chi^2/$dof), in scheme A. 
We present the distribution of the sum of the these $\hat{\chi}^2$ with respect to some fixed $g_{\rho\pi\pi}$ and $\Lambda_{\rho\pi\pi}$ in Fig.~\ref{fig:fixgL}.  
It is straightforward to find the preferred values of the coupling and cutoff from this distribution. 
It leads us to choose the values at the black point, where $g_{\rho\pi\pi}=7.07$ 
and $\Lambda_{\rho\pi\pi}=890$ MeV for $N_f=2+1$. 
On the other hand, it is found that for $N_f=2$, we should use slightly different values, namely $g_{\rho\pi\pi}=6.75$ and $\Lambda_{\rho\pi\pi}=950\,$MeV~\cite{Hu:2017wli, Hu:2016shf}.
These values ensure the $\hat{\chi}^2$ satisfies the condition $\hat{\chi}^2\lesssim 2$ for all 
of these values of
$m_\pi$, as illustrated in Table~\ref{tab:Fit Results Table}, in the column of $\hat{\chi}^2$ for scheme B.
\begin{figure}[tbp!]
    \centering
    \includegraphics[width=.5\textwidth]{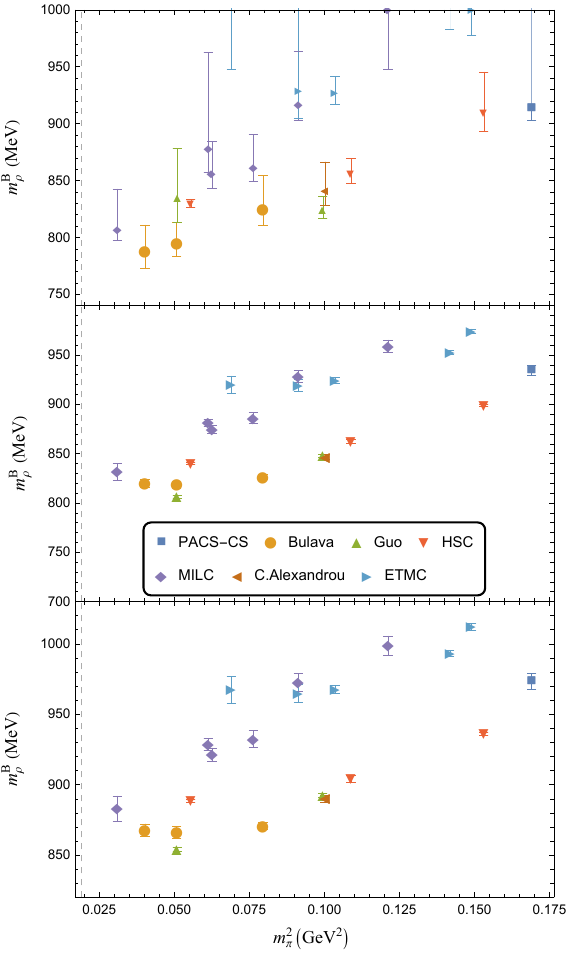}
    \caption{
    $m_\pi$-dependence of $m_\rho^\B$ in schemes A, B and C from top to bottom. The dashed gray vertical line indicates the physical $m_\pi$ value.
    }
\label{Fig:Result-m}
\end{figure}

For scheme B, the curves denoting the spectra as a function of the spatial extent, $L$, are shown as orange dot-dashed lines in the figures in Appendix~\ref{app:fiting levels}.
The fitted $m_\rho^\B$ values are presented in Fig.~\ref{Fig:Result-m}, and the columns for scheme B in Table~\ref{tab:Fit Results Table}.
Comparing Fig.~\ref{Fig:Result-m}(A) with (B), one finds that the uncertainties in $m_\rho^\B$ are significantly reduced, albeit with slight shifts in the central values.
The $m_\pi$-dependence of $m_\rho^\B$ remains unclear.
In particular, the values of $m_\rho^\B$ at $m_\pi=301$, $302$, $315$, $316$ and $322$ MeV extracted from MILC, ETMC, Guo \textit{et al}., C.~Alexandru \textit{et al}., and ETMC respectively, significantly fluctuate between $840$ and $920$ MeV.
It is apparent that there are substantial differences in the results from different LQCD groups.
To address this issue, we examine the contribution of the coupling to the 
$\omega\pi$~\cite{Bruns:2004tj, Leinweber:2001ac} channel, which is the closest two-particle channel in the present work and which, significantly, yields the leading non-analytic contribution to the $\rho$ self-energy.

\subsubsection{Results for Scheme C}

Once $V_{\omega\pi}$ is turned on, two additional parameters, $g_{\omega\rho\pi}$ and $\Lambda_{\omega\rho\pi}$, are involved.
In addition, $m_\omega$ also depends on $m_\pi$.
%
The $m_\omega$ is not provided by most of the lattice QCD collaborations considered here.
However, HSC reported a value of $m_\omega$ at $m_{\pi} = 391$ MeV~\cite{Dudek:2012xn}.
It provides an excellent opportunity to quantify the importance of quark-mass changes.
They find a $\rho$-meson resonance position of 
$a_t\,m_R     = 0.15085(18)(3)$ in the Briet-Wigner parameterization
to be compared with their stable omega mass of $a_t\,m_{\omega} = 0.15678(41)$.  
The difference/average is only $3.9\%$ even at this very large pion mass.
Since modern lattice QCD results are at lower quark masses, in our calculation for other collaborations, we just use the approximation $m_\omega = m_\rho^{\text{BW}}$.

Since the threshold of $\omega\pi$ is higher than the spectrum extracted, it is expected that the $\omega\pi$ loop will shift the value of $m_\rho^\B$ but have a negligible influence on the resulting $\chi^2$.

Instead of allowing $g_{\omega\rho\pi}$ and $\Lambda_{\omega\rho\pi}$ to be two 
additional free parameters, we impose appropriate constraints to fix them.
Two constraints are identified at the physical pion mass $\mu_\pi$: the decay width,  
$\Gamma_{\omega\to 3\pi}$, is primarily determined by the $\omega\to\rho\pi\to 3\pi$ mechanism (which is estimated to yield around $90\%$~\cite{Kleefeld:2001xd} of the width) and the $P$-wave phase shifts of $\pi\pi \to \pi\pi$ in the energy region around the $\rho$ mass.

For simplification, we take the two cut-off, $\Lambda_{\rho\pi\pi}$ and 
$\Lambda_{\omega\rho\pi}$, to share the same value.
Consequently, there are four undetermined parameters left, namely two coupling constants, one cut-off and one bare mass.
Once $g_{\rho\pi\pi}$ and $\Lambda_{\rho\pi\pi}$ are fixed, the other parameters at $\mu_\pi$ can also be determined from the decay width of $\omega\to 3\pi$ and the $P$-wave phase shifts of $\pi\pi \to \pi\pi$ at the physical pion mass $\mu_\pi$.

The detailed procedure for parameter determination is presented in Appendix~\ref{app:gcutomega}.
Here, we simply summarize the preferred values:  $g_{\rho\pi\pi}=7.07\,(6.75)$ and 
$\Lambda_{\omega\rho\pi}=\Lambda_{\rho\pi\pi}=900\,(980)$ MeV for $N_f=2+1$ ($N_f=2$) and  $g_{\omega\rho\pi}=18/$GeV for either $N_f$.
It is worth mentioning that our value of $g_{\omega\rho\pi}$ is similar to that used in Ref.~\cite{Leinweber:2001ac}, $16/$GeV. 
Additionally, $g_{\rho\pi\pi}$ and $\Lambda_{\omega\rho\pi}$ are slightly shifted compared to those found using scheme B, because of the introduction of the $\omega\pi$ channel. 
We then proceed to minimize the total $\chi^2$ by fitting the value of $m_\rho^\B$ for each pion mass.

The fitted spectra as functions of the spatial extent, $L$, for scheme C are shown as black dotted lines in the figures of Appendix~\ref{app:fiting levels}, which illustrates only minor discrepancies from the orange dot-dashed lines found using scheme B, as expected.
The fitted values of $m_\rho^\B$ are presented in the lower portion of Fig.~\ref{Fig:Result-m} and are tabulated in the columns for scheme C in Table~\ref{tab:Fit Results Table}. 
The preferred values of $m_\rho^\B$ are about $50$ MeV higher than found in scheme B because of the additional self-energy term, $\Sigma_{\omega\pi}$, as defined 
in Eq.~(\ref{eq:self omegapi}).

Even after including the effect of the $\omega\pi$ coupled channel, the $m_\pi$-dependence of $m_\rho^\B$ shown in Fig.\ref{Fig:Result-m} is still scattered.
It becomes apparent that the bare masses extracted from different lattice groups do not permit a consistent interpretation, which indicates the presence of intrinsic systematic differences between the lattice spectra provided by different collaborations.
Such discrepancies lead us to consider the following issues that may influence the lattice results presented. These include:
\begin{itemize}
    \item Different residual lattice artefacts due to the different gauge and fermion actions considered.
    \item Varied scale-setting schemes employed by different collaborations.
    \item Different methods used to extract the finite volume spectra.
\end{itemize}
In the absence of systematic errors, the results provided by different collaborations should be consistent with each other after finite volume and lattice spacing artefacts are taken into account.
Our HEFT approach enables one to account for the finite volume of the lattice, several values of the pion mass, as well as lattice spacing artefacts, all within a single formalism. 
As such, this is the first examination of the self consistency of world lattice QCD results for $\pi\pi$ scattering in $\rho$ meson channel.

\subsection{Extrapolation in $m_\pi$}
\label{sec:extrap}
In the last section, we present the outcomes of our fitting approach applied to the finite volume spectra provided by various collaborations for a wide range of values of $m_\pi$.
With these results, we now investigate the $m_\pi$-dependence of the properties of the $\rho$ meson and extrapolate them into the physical region.
In scheme A, our investigations reveal that both $g_{\rho\pi\pi}$ and $\Lambda_{\rho\pi\pi}$ display little variation as $m_\pi$ varies. 
In the spirit of chiral effective field theory, the couplings and regulator parameters are held fixed.
Thus, we could concentrate on the $m_\pi$-dependence of the bare $\rho$ mass, $m_\rho^\B$,  using Eq.~(\ref{eq:mpi depdence of mrhoB}).
It is possible to extrapolate the fitting results of $m_\rho^\B$ in schemes B and C, but not in scheme A, since there the values of $m_\rho^\B$ are correlated with $g_{\rho\pi\pi}$ and $\Lambda_{\rho\pi\pi}$. 

In principle, it is natural to consider putting all the values of $m_\rho^\B(m_\pi)$ together and performing a global fit to make full use of the lattice data. 
However, from the two lower figures in Fig.~\ref{Fig:Result-m}, it is hard to extract useful information, since the data show large inconsistent variations. 
The possible reasons have been discussed in the previous section.

For example, with reference to the discussion about the lattice spacing effect in 
Eq.~(\ref{eq:mpi depdence of mrhoB with lat spacing}), the coefficient $\xi$ are different for each LQCD group in principle.
%
To confirm this, we applied Eq. (41) with a single value of $\xi$ and
found a large $\chi^2$. 

Also, there is a huge difference between the values of observables calculated using the parameters extrapolated to physical $m_\pi$ and those measured in experiments.
This suggests that we should make the extrapolation of the data to the physical point collaboration by collaboration.

\begin{figure}[tbp]
    \centering
    \includegraphics[width=.38\textwidth]{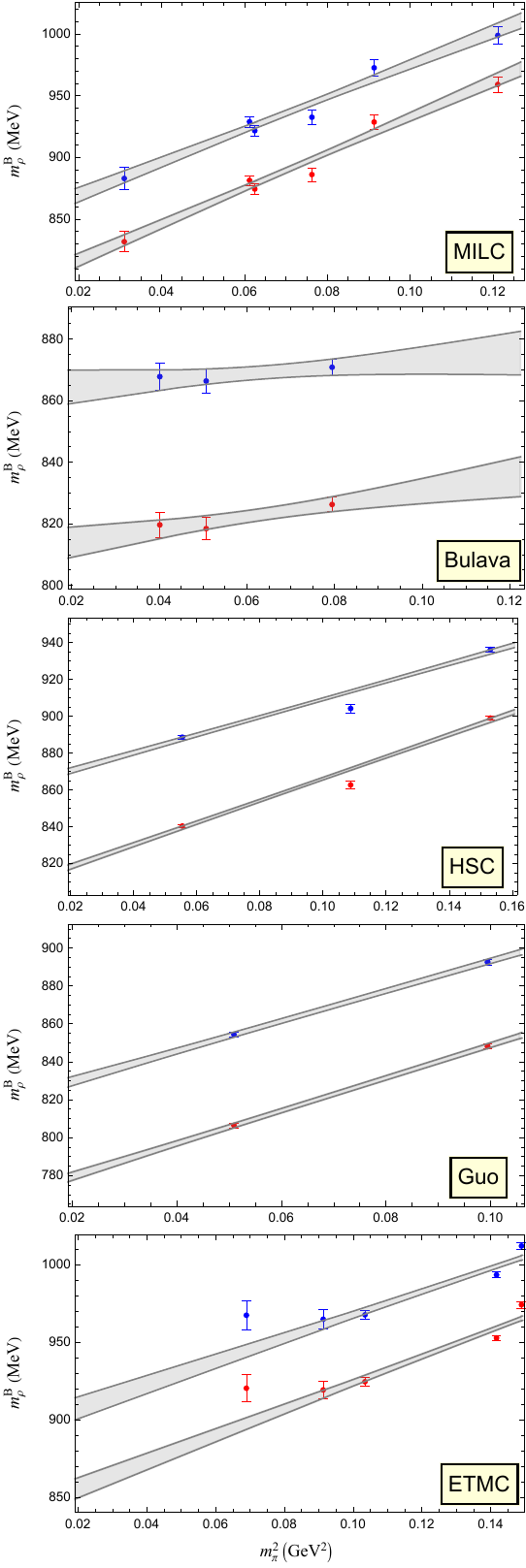}
    \caption{
    $m_\pi$-dependence and extrapolation of $m_\rho^\B$ for each collaboration. Red and blue points indicate the fitting results of $m_\rho^\B$ in schemes B and C, respectively. Gray bands represent the quadratic function $m_\rho^\B = c_0 + c_1\, m_\pi^2$ with uncertainty, where $c_0$ and $c_1$ for both schemes are given in Table~\ref{Tab:extrapolation}. For convenience the lower bound of the $m_\pi^2$-axis is set as the physical value $\mu_\pi^2$. 
    }
\label{Fig:extrapolation}
\end{figure}

Because for each group, there are only a limited number of values of $m_\pi$ and the lattice spacing does not change a lot, the lattice spacing term can be absorbed and we just use Eq.~(\ref{eq:mpi depdence of mrhoB}) to perform the extrapolations.
Furthermore, we have two free parameters in Eq.~(\ref{eq:mpi depdence of mrhoB}), thus only the data of collaborations having no less than two $m_\pi$ points are analyzed.
The fitting and extrapolation results are shown in Fig.~\ref{Fig:extrapolation} and Table~\ref{Tab:extrapolation}. 
\begin{table}
 \caption{Extrapolation results are summarized for schemes B and C. For each collaboration, the results of B and C are given in the first and second row, respectively. The second and third column present the coefficient $c_0$ and $c_1$ defined in Eq.(\ref{eq:mpi depdence of mrhoB}). The fourth column presents the extrapolated $m_\rho^\B$ at the physical pion mass. The fifth column presents the pole mass, defined by Eq.(\ref{eq:pole of propagator}). 
    }
    \centering
    \begin{tabular}{ccccccccc}
    \toprule
    {COLLAB.} & $c_0$(MeV) & $c_1(\mathrm{GeV}^{-1})$ & $m_\rho^\B(\mu_\pi)$ & $m_\rho^\p(\mu_\pi)$  \\
    \midrule
    \multirow{2}{*}{Bulava} & 809.8(7.0) & 0.21(0.11) &  814.0(5.0) & 765.0(6.0) \\
    \cmidrule{2-5}
    & 862.3(7.6) & 0.11(0.12) & 864.0(6.0) & 765.0(6.0) \\ 
    \midrule
    \multirow{2}{*}{MILC} & 788.0(7.3) & 1.45(0.10) & 816.0(6.0) & 768.0(6.0) 
    \\
    \cmidrule{2-5}
    & 843.3(7.95) & 1.32(0.11) & 869.0(6.0) & 769.0(6.0) \\
    \midrule
    \multirow{2}{*}{HSC} & 806.7(1.71) & 0.60(0.02) & 818.2(1.4) & 770.6(1.7) \\
    \cmidrule{2-5}
    & 861.3(1.9) & 0.49(0.02) & 870.7(1.6) & 771.3(1.7)  \\
    \midrule 
    \multirow{2}{*}{ETMC} & 838.9(7.7) & 0.85(0.06) & 855.0(7.0) & 814.0(8.0)
    \\
    \cmidrule{2-5}
    & 892.5(8.3) & 0.75(0.06) & 907.0(7.0) & 809.0(7.0) \\
    \midrule
    \multirow{2}{*}{Guo} & 762.2(2.9) & 0.86(0.04) & 778.8(2.3) & 719.3(2.6)
    \\
    \cmidrule{2-5}
    & 813.8(3.2) & 0.79(0.04) & 829.0(2.5) & 719.0(2.6) \\ 
    \bottomrule
    \end{tabular}
\label{Tab:extrapolation}
\end{table}

As shown in Fig.~\ref{Fig:extrapolation}, there are five collaborations having no less than two different $m_\pi$ points.
For each collaboration, the points show a good linear relation between $m_\rho^\B$ and $m^2_\pi$, whether the $\omega\pi$ loop is included or not. 
The only notable exception is one point with a large uncertainty from ETMC. 
With $c_0$ and $c_1$ determined, we can obtain the bare mass of the $\rho$ at the physical pion mass. 
Subsequently, we can get $m_\rho^\p$ by solving Eq.~(\ref{eq:pole of propagator}).
The results are listed in the last column of Table~\ref{Tab:extrapolation}.

For MILC, HSC and Bulava \textit{et al}., even though their values of $c_0$ and $c_1$ are quite different, the extrapolated $m_\rho^\p$ all agree with the experimental value.
However, for ETMC and Guo \textit{et al}., they are about $30$ MeV higher and $50$ MeV lower compared to the experimental value, respectively. 
The relatively high $m_\rho^\p$ obtained by ETMC is not so surprising, since in their previous work~\cite{Feng:2010es} a higher value of $m_\rho$ compared to the others was also reported.
The lower $m_\rho^\p$ extracted from Guo \textit{et al}. also agrees with their own result,  presented in Ref.~\cite{Guo:2016zos}, which possibly results from using $N_f=2$.
Clearly, the physical $\rho$ masses obtained here all indicate the consistency between our method and previous work, while in addition we provide detailed information on the 
$m_\pi$-dependence of $m_\rho^\B$. 
\begin{figure}[tbp]
    \centering
    \includegraphics[width=.5\textwidth]{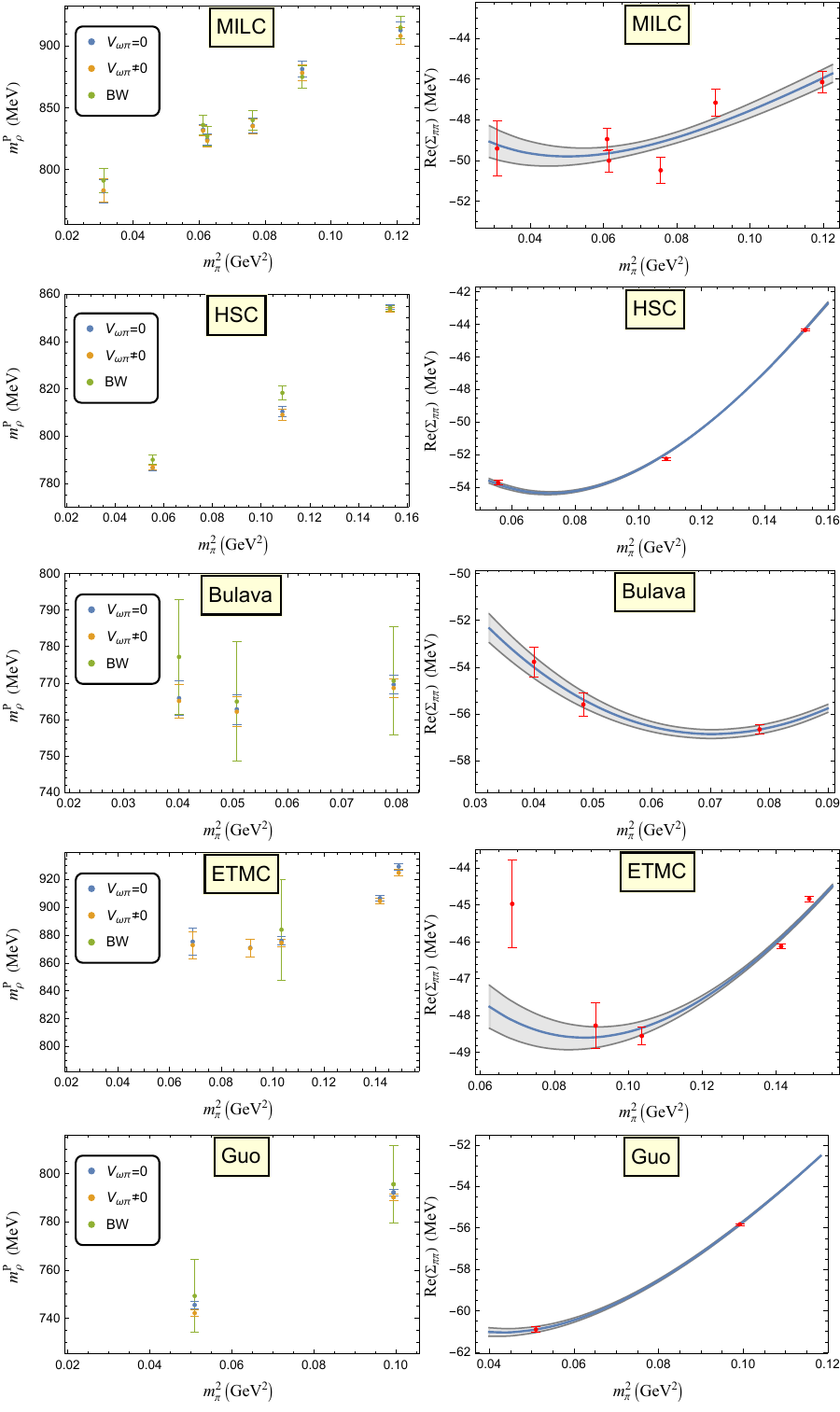}
    \caption{
    $m_\pi$-dependence of $m_\rho^\p$ and the real part of the self-energy $\Sigma_{\pi\pi}$. Left column: $m_\rho^\p$ defined by Eq.(\ref{eq:pole of propagator}) using parameters for schemes B and C, and $m_\rho$ obtained from the Breit-Wigner parameterization of the phase shift provided in each paper. Right column: real part of the self-energy $\Sigma_{\pi\pi}$. Blue lines with gray bands represent the $\Re\Sigma_{\pi\pi}= m_\rho^\p(m_\rho^\B) - m_\rho^\B$ as in Eq.(\ref{eq:mrho approxiamte condition HEFT}) using  $m_\rho^\B = c_0 + c_1 \,m_\pi^2$ with $c_0$ and $c_1$ being the extrapolation coefficients for scheme B presented in Table \ref{Tab:extrapolation}. Red points also represent $\Re\Sigma_{\pi\pi}$ but using $m_\rho^\B$ from fitting results of scheme B presented in Table~\ref{tab:Fit Results Table}.
    }
\label{fig:PoleMassAndSE}
\end{figure}
%

\subsection{Discussion and Exploration}

In this section, we make some remarks concerning the numerical results.
In scheme A, we first found that the coupling constants $g_{\rho\pi\pi}$ and cut-off 
$\Lambda_{\rho\pi\pi}$ are both weakly dependent on $m_\pi$.
This conclusion supports our previous study of the baryon resonances, $\Lambda(1405)$, $N^*(1535)$ and $N^*(1440)$ (Roper), where we only considered the $m_\pi$ dependence of the masses of various hadrons but not the couplings and cut-off.
Furthermore, the large uncertainty found for the cut-off is also acceptable, because in principle the physical observables should not be so sensitive to it.

In Eq.~(\ref{eq:mpi dependence of mrho perturbation theory}), there are many nonlinear terms in $m_\pi^2$, arising from the self-energy part.
In the right column of Fig.~\ref{fig:PoleMassAndSE}, we show the value of the 
$\Re\Sigma_{\pi\pi}$ at the pole position as a function of $m_\pi^2$.
Given $m_\rho^\B$, $\Re\Sigma_{\pi\pi}$ is calculated by $m_\rho^\p-m_\rho^\B$ as in Eq.~(\ref{eq:mrho approxiamte condition HEFT}), where $m_\rho^\p$ is obtained from Eq.~(\ref{eq:pole of propagator}).
Scalars in the figure are calculated by using the values of $m_\rho^\B$ from the fitting results of scheme B presented in Table~\ref{tab:Fit Results Table}, while the lines with error bands use $m_\rho^\B=c_0+c_1 \, m_\pi^2$, with $c_0$ and $c_1$ being the coefficients for scheme B presented in Table \ref{Tab:extrapolation}.

Clearly both these scalars and lines in Fig.~\ref{fig:PoleMassAndSE} exhibit a nonlinear behavior. 
However, the difference of $\Re\Sigma_{\pi\pi}$ at different $m_\pi$ are just around 10 MeV when $m_\pi$ varies over the range $140$ to $400$ MeV.
Therefore, there will also be an approximately linear relation between $m_\rho^\p$ and $m_\pi^2$, as shown in the left column of Fig.~\ref{fig:PoleMassAndSE}, where the Breit-Wigner masses provided by the three collaborations are also shown. 
This is the reason why the Breit-Wigner $\rho$ mass could be described well by a linear function of $m_\pi^2$ in Ref.~\cite{Fu:2016itp}.

Things are similar even when the $\omega\pi$ channel is introduced.
From Table~\ref{Tab:extrapolation} one can see that $m_\rho^\p$ is nearly unchanged while $m_\rho^\B$ changes a lot when $V_{\omega\pi}$ turns on. 
Nonetheless, $m_\rho^\B$ still exhibits a linear relation to $m_\pi^2$ as shown in 
Fig.~\ref{Fig:extrapolation}, where the two lines are almost parallel for each LQCD group.
Also, it is apparent that the contribution from the $\omega\pi$ loop only makes a significant change in the value of $c_0$, but just slightly modifies the slope, $c_1$.
These facts suggest that $\Re\Sigma_{\omega\pi}$ is also weakly dependent on $m_\pi$ and its effect can be effectively absorbed into $c_0$.

In summary, the slow variation of the contributions from $\pi\pi$ and $\omega\pi$ loops significantly affects $c_0$, while only slightly influencing the slope, $c_1$.
Therefore, it makes little sense to talk about $m_\rho^\B$ solely based on experimental results, as the fitted $m_\rho^\B$ strongly depends on how the hadronic loops are estimated.
It is the slope $c_1$ that contains more useful, less model dependent, physical information concerning the structure of the $\rho$ meson, which can only be extracted from the LQCD data at unphysical $m_\pi$.
Furthermore, in principle, on the theoretical side the slope $c_1$ can be calculated at the quark-level in various models.
Thus, with the help of $c_1$ the relevant models could be distinguished.
This is quite a good example of the idea that the data extracted at unphysical values of $m_\pi$ are able to provide us with additional information concerning the structure of hadrons.

The linear relation between $m_\rho^\B$ and $m_\pi^2$ is consistent with the assumption that $\ket{\rho_\B}$ is a pure $q\bar{q}$ state.
Additionally, the contribution from hadron loops to $m_\rho^\p$ accounts for only approximately 20\% of the total mass for the optimal value of the regulator parameter.
Consequently, we can also conclude that the bare $\rho$ plays the most important role in the structure of the observed $\rho$ meson.
To confirm this, we pick out several eigenstates whose energy is close to the physical $\rho$ mass and look at their composition, a feature which is characteristic of HEFT.
The eigenstates of the FVH are the counterpart in the finite volume of continuous scattering states in the infinite volume~\cite{Wu:2017qve}.
Therefore, obviously, it is expected that the eigenstate whose energy is closest to the resonance region is the counterpart of the $\rho$.
For illustration, we pick five eigenstates from the spectrum generated by HSC(2015) with $m_\pi=236$ MeV and calculate the probabilities of the bare state, $\pi\pi$ and $\omega\pi$ components. 
The results are presented in Table~\ref{tab:composition of state}.
It is found that the component $\rho_\B$ has a probability around 75\%. That is, it is definitely dominant.
\begin{table}[t]
    \caption{Composition of eigenstates $\ket{\psi}$ of some energy levels from HSC(2015). In the first column, $\ket{\psi}$ is labelled by $(\boldsymbol{n}^2,\Gamma)$, as it is the eigenstate that has the largest $\rho_\B$ component among those whose energies are extracted by the operators in representation $\Gamma$ and with total momentum $\boldsymbol{P}^2=(\frac{2\pi}{L})^2\boldsymbol{n}^2$, as shown in Fig.~\ref{fig:spectrum curve of L}. In the subsequent column we present the composition of $\ket{\psi}$, 
i.e, $\abs{\braket{\phi}{\psi}}^2$ with $\ket{\phi}=\ket{\rho_\B},\ket{\pi\pi}$ for $V_{\omega\pi}=0$ and also $\ket{\omega\pi}$ for $V_{\omega\pi}\neq0$.
    }    
    \begin{tabular}{ccc@{\hspace{5pt}}|@{\hspace{5pt}}ccc}
    \toprule
        \multirow{2}{*}{$(\boldsymbol{n}^2,\Gamma)$} & \multicolumn{2}{c}{$V_{\omega\pi}=0$} & \multicolumn{3}{c}{$V_{\omega\pi}\neq 0$}
        \\
        \cmidrule(lr){2-3} \cmidrule(lr){4-6} 
         & $\rho_\B$ & $\pi\pi$ 
         & $\rho_\B$ & $\pi\pi$ &  $\omega\pi$ \\
        \midrule
        $(1,A_1)$ & 0.7365 & 0.2635 & 0.6966 & 0.2664 & 0.0370
        \\
        \midrule
        $(2,A_1)$ & 0.7963 & 0.2037 & 0.7537 & 0.2028 & 0.0434
        \\
        \midrule
        $(3,A_1)$ & 0.7701 & 0.2299 & 0.7295 & 0.2241 & 0.0464
        \\
        \midrule
        $(4.A_1)$ & 0.7432 & 0.2568 & 0.7093 & 0.2503 & 0.0404
        \\
        \midrule
        $(3,E)$ & 0.6514 & 0.3486 & 0.6171 & 0.3394 & 0.0434
        \\ 
        \bottomrule
    \end{tabular}
\label{tab:composition of state}
\end{table}

\begin{figure}[t]
    \centering
    \includegraphics[width=.45\textwidth]{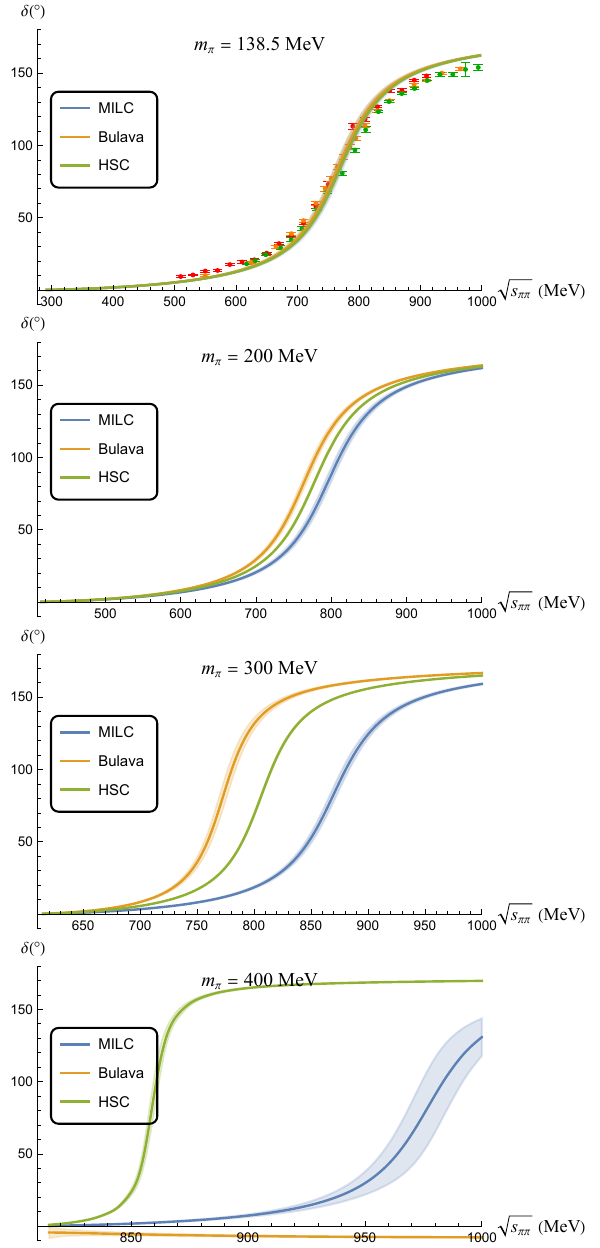}
    \caption{
    Phase shift calculated by Eq.~(\ref{eq:phaseshift}) using the parameters $g_{\omega\rho\pi}=0$, $g_{\rho\pi\pi}=7.07$, $\Lambda_{\rho\pi\pi}=890\,$MeV and $m_\rho=c_0 + c_1 m_\pi^2$ with $c_0$ and $c_1$ being the extrapolation coefficients for scheme B given in Table~\ref{Tab:extrapolation}. The points in the first figure are experimental values from Ref.~\cite{Protopopescu:1973sh, Estabrooks:1974vu, Hyams:1973zf}. The line for Bulava at $m_\pi=400$ MeV denotes a $\pi\pi$ bound state.
    }
\label{fig:PhaseShift}
\end{figure}

With the parameters listed in Table~\ref{tab:Fit Results Table}, we can predict the $\pi\pi \to \pi\pi$ $P$-wave phase shift at unphysical masses.
Here we focus on the three collaborations whose values of $m_\rho^\p$ at the physical $m_\pi$ agree with the experimental value.
In Fig.~\ref{fig:PhaseShift}, we show the phase shifts calculated for four different values of $m_\pi$.
In the first row, at the physical point, the phase shifts of the three collaborations are nearly the same. 
This is expected since their values of $m_\rho^\B$ at physical $m_\pi$ are consistent.
They are in good agreement with the experimental data, except for the region away from resonance, where the $\pi\pi-\pi\pi$ $t/u$-channel interaction may not be negligible.

In other rows we predict the phase shifts of the three collaborations at three unphysical values of $m_\pi$. 
Since the values of $c_0$ and $c_1$ obtained by analysis of the data from these three collaborations are very different, their phase shifts are not consistent.
Especially, for $m_\pi=400$ MeV, 
we predict that the typical line shape would disappear in the phase shifts of $\pi\pi$ scattering for Bulava \textit{et al}., since it would be a bound state of $\pi\pi$.

The most serious problem found in this work is that the two coefficients, $c_0$ and $c_1$, differ a lot for different LQCD groups.
This unexpected variation arises from differences in the lattice QCD simulations, the interpolating fields considered in constructing correlation matrices, the analysis methods applied to the correlation matrices and finally the scale-setting schemes.

The use of different $\mathcal{O}(a)$-improved actions gives rise to different $\mathcal{O}(a^2)$ errors such that the coefficient of $a^2$ differs for each set of lattice QCD results. 
It will be important to have two or more different lattice spacings available in high-quality sets to enable a determination and elimination of this lattice artefact.

State isolation is key to measuring the subtle shifts of the finite-volume energies from their non-interacting energies, vital to measuring a phase shift. 
In obtaining consistency across lattice collaborations, modern projected correlator methods should be adopted to reduce excited state contamination in the extracted energy eigenvalues.
Moreover, all nonlocal two particle momentum-projected interpolating fields participating in the resonance region need to be considered and mixed with the single particle operators to ensure multiple eigenstates are not participating in the state-projected correlation functions of the correlation matrix. 
In the present case, single particle $\rho$ interpolations are to be mixed, not only with $\pi\pi$ correlations, but also with $\pi\omega$ correlators. 
Indeed, Fig.~\ref{fig: spectra example} and the figures in Appendix.\ref{app:fiting levels}, illustrating the fits of HEFT to contemporary lattice QCD results, raise concerns about multi-state contamination in the analysis.

Finally, the choice of scale setting scheme is of vital importance when attempting to describe QCD properties away from the physical point. 
While all schemes are designed to extrapolate to the physical point, the manner in which they move away from the physical point is different. 
The Sommer scale is designed to be physical. 
Maintaining physics at the scale relevant to the charmonium spectrum, it naturally includes changes in the renormalization of the strong coupling constant due to changes in the sea quark masses. 
However, other schemes where the quark mass has no effect on the coupling constant are possible, provided the only goal is to get to the physical point. 
In light of the plethora of scale setting schemes currently proposed, we encourage the consideration of how well a proposed scheme is suited to learning the properties of QCD in a universe with different quark masses.

\section{Summary and Outlook}
\label{sec:summary}

In this work, we collected finite volume spectra for the $I=\ell=1$ $\pi\pi$ sector provided by LQCD collaborations over the past decade. 
These spectra were fit in a consistent manner within the framework of Hamiltonian Effective field theory. 
The basic states included in the Hamiltonian were a bare $\rho$ state and the $\pi\pi$ and $\omega\pi$ coupled channels.
In this framework, we successfully fit the finite volume spectra in the rest frame, moving frame and the elongated box, and complemented this with experimental data for the $\pi\pi \to \pi\pi$ $P$-wave phase shifts. 

We employed three schemes to fit the energy levels obtained at various pion masses.
Through scheme A, we found that $g_{\rho\pi\pi}$ and 
$\Lambda_{\rho\pi\pi}$ exhibit a weak dependence on $m_\pi$. 
In the following scheme B, we set these two parameters as constant in order to obtain the $m_\pi$-dependence of the bare $\rho$ basis state, $m_\rho^\B(m_\pi)$.
In scheme C, where the $\omega\rho\pi$ vertex was included with additional constraints from experimental data, we again extracted $m_\rho^\B(m_\pi)$. 
Finally, we used the linear relation between $m_\rho^\B$ and $m_\pi^2$, as shown in 
Eq.~(\ref{eq:mpi dependence of mrho perturbation theory}), to perform an extrapolation.
Because the relationship between $m_\rho^\B$ and $m_\pi^2$ was highly dependent on the LQCD group whose data we used, we were unable to fit all $m_\rho^\B$ simultaneously, and resorted to extrapolating collaboration by collaboration. 

Based upon the extrapolations of data from the five LQCD groups, it was found that for each collaboration, $m_\rho^\B(m_\pi)$ could be described well by Eq.~(\ref{eq:mpi depdence of mrhoB}). 
This supports the hypothesis that the single-particle bare $\rho$ meson plays an important role in forming a physical peak in the $P$-wave $\pi\pi$ scattering.
The pole mass $m_\rho^\p$ defined by Eq.~(\ref{eq:mrho approxiamte condition HEFT}) are also calculated. When extrapolated to physical pion mass $\mu_\pi$, the $m_\rho^\p$ for MILC, Bulava {\it et al.} and HSC agree with the experimental measurements. We can also predict the observables such as the phase shift for each collaboration at some certain unphysical $m_\pi$.  
For each group, the pole mass $m_\rho^\p$ and the coefficients $c_1$ in 
Eq.~(\ref{eq:mpi depdence of mrhoB}) were stable with and without the $\omega\pi$ loop contribution, while the bare mass had a shift of around $50$ MeV.
%
%
Unfortunately, from the current LQCD data, the extracted value of $c_1$ is dependent on the lattice collaboration whose data is used. 
This indicates the important discrepancies in the results of today's lattice QCD calculation among different collaborations due to the possible reasons discussed previously.
In the future it would obviously be ideal to have lattice data for the $\rho$ meson at different pion masses with all systematic effects carefully controlled in order to resolve the $m_\pi$-dependence of the $\rho$ meson.

\begin{acknowledgments}
We acknowledge useful discussions and valuable comments from 
Xu Feng, Feng-Kun Guo, Chuan Liu,  Liuming Liu, Guang-Juan Wang, Mao-Jun Yan, Yi-Bo Yang, Zhi Yang, Bing-Song Zou. 
This work was partly supported  by the National Natural Science Foundation of China (NSFC) under Grants Nos. 12175239 and 12221005 (J.J.W), and by the National Key R$\&$D Program of China under Contract No. 2020YFA0406400 (J.J.W), and by the Chinese Academy of Sciences under Grant No. YSBR-101 (J.J.W).
This research was also undertaken with assistance of resources from the National Computational Infrastructure(NCI), provided through the National Computational Merit Allocation Scheme. This work was supported by the Australian Research Council through Discovery Projects DP190102215 and DP210103706 (DBL) and DP230101791 (AWT).
Y. L. is supported by the Excellence Hub project "Unraveling the 3D parton structure of the nucleon with lattice QCD (3D-nucleon)" id EXCELLENCE/0421/0043 co-financed by the European Regional Development Fund and the Republic of Cyprus through the Research and Innovation Foundation. Y. L. further acknowledges computing time granted on Piz Daint at Centro Svizzero di Calcolo Scientifico (CSCS) via the project with id s1174, JUWELS Booster at the J\"{u}lich Supercomputing Centre (JSC) via the project with id pines, and Cyclone at the Cyprus institute (CYI) via the project with ids P061, P146 and pro22a10951.

\end{acknowledgments}

\bigskip

\bibliography{ref}

\clearpage

\appendix
\section{The formulas for $\boldsymbol{n}\leftrightarrow\boldsymbol{k}^*(\boldsymbol{n})$ and $e_{\boldsymbol{n}}$ }
\label{app:knen}

Here, we will list all cases involved in our present calculation, since the transformation $\boldsymbol{n}\leftrightarrow\boldsymbol{k}^*(\boldsymbol{n})$ and $e_{\boldsymbol{n}}$ differ from how the momentum is quantized. 
A summary is given here. 
\begin{enumerate}
    \item If quantized in a cubic box with total momentum $\boldsymbol{P}=0$ and length $L$, then $\boldsymbol{k}^*(\boldsymbol{n})=\frac{2\pi}{L}\boldsymbol{n}$, and $e_{\boldsymbol{n}}=\boldsymbol{n}^2$.
    \item If quantized in a elongated box with total momentum $\boldsymbol{P}=0$, then $\boldsymbol{k^*}(\boldsymbol{n}) = \frac{2\pi}{L}\left(\boldsymbol{n}_\bot + \frac{1}{\eta}\boldsymbol{n}_\parallel \right)$ where $\eta>1$ denotes the elongation strength, $\bot$ and $\parallel$ denotes the vertical and parallel component to the elongated direction named $\boldsymbol{d}$, respectively.
    Then $e_{\boldsymbol{n}}$ need two values, $\boldsymbol{n}^2$ and $|\boldsymbol{n}\cdot\boldsymbol{d}|$.
    \item If quantized in a cubic box with total momentum $\boldsymbol{P}\equiv2\pi\boldsymbol{d}/L\neq 0$, then $\boldsymbol{k}^*(\boldsymbol{n}) = \boldsymbol{k}^*(\boldsymbol{k}(\boldsymbol{n}))$ and $\boldsymbol{k}(\boldsymbol{n})=\frac{2\pi}{L}\boldsymbol{n}$ where $\boldsymbol{k}^*(\boldsymbol{k})$ is the ``Lorentz-like transformation" for channel $\alpha$ defined as
\begin{align}\label{eq:mmt}
    &\boldsymbol{k}^*(\boldsymbol{k}) = \boldsymbol{k}_{\perp} + \gamma \left( \boldsymbol{k}_{\parallel} - \frac{E_{\alpha_1}(\boldsymbol{k})}{E_{\alpha_1}(\boldsymbol{k})+E_{\alpha_2}(\boldsymbol{P}-\boldsymbol{k})} \boldsymbol{P} \right) \,, \nonumber\\
    &\gamma = \frac{E_{\alpha_1}(\boldsymbol{k})+E_{\alpha_2}(\boldsymbol{P}-\boldsymbol{k})}{\sqrt{\left( E_{\alpha_1}(\boldsymbol{k})+E_{\alpha_2}(\boldsymbol{P}-\boldsymbol{k}) \right)^2-\boldsymbol{P}^2}} \,,
\end{align}
where $\boldsymbol{k}_{\parallel}=(\boldsymbol{k}\cdot\boldsymbol{P})\boldsymbol{P}/\boldsymbol{P}^2$ and $\boldsymbol{k}_{\perp}=\boldsymbol{k}-\boldsymbol{k}_{\parallel}$. 
Correspondingly the Jacobian is
\begin{align}\label{eq:jacob}
    &\boldsymbol{J}(\boldsymbol{k})=\frac{E_{\alpha_1}(\boldsymbol{k})+E_{\alpha_2}(\boldsymbol{P}-\boldsymbol{k})}{E_{\alpha_1}(\boldsymbol{k})\,E_{\alpha_2}(\boldsymbol{P}-\boldsymbol{k})}\bigg{/}\frac{E_{\alpha_1}(\boldsymbol{k}^*)+E_{\alpha_2}(\boldsymbol{k}^*)}{E_{\alpha_1}(\boldsymbol{k}^*)\,E_{\alpha_2}(\boldsymbol{k}^*)} \,.
\end{align}
And the $e_{\boldsymbol{n}}$ also needs two values, $\boldsymbol{n}^2$ and $\boldsymbol{n}\cdot\boldsymbol{d}$.
Furthermore, please note that for the $\pi\pi$ case, the two values are unordered.
    \item If quantized in a elongated box with total momentum $\boldsymbol{P}\neq 0$ and $\boldsymbol{P}$ is parallel to the elongated direction $\boldsymbol{d}$, it is same as the third case above except that $\boldsymbol{k}(\boldsymbol{n}) = \frac{2\pi}{L}\left(\boldsymbol{n}_\bot + \frac{1}{\eta}\boldsymbol{n}_\parallel \right)$
And the $e_{\boldsymbol{n}}$ are also the same as that in the third case.
\end{enumerate}

\section{Determination of fixed parameters in Scheme C}
\label{app:gcutomega}

In this appendix the determination of $g_{\omega\rho\pi}$ and $\Lambda_{\omega\rho\pi}$ are discussed. As mentioned in the main text, besides lattice spectrum the parameters should also be constrained by the decay width $\Gamma_{\omega\to 3\pi}$ as well as phase shift $\delta_{\pi\pi\to\pi\pi}^{\ell=1}$ on the experimental side.

The decay channel $\omega\to 3\pi$ is believed to be dynamically dominated by the $\omega\to\rho\pi\to 3\pi$ mechanism. Therefore, the calculation of the decay width would concern $V_{\pi\pi}$ and $V_{\omega\pi}$ defined in Eq.(\ref{eq:inf Vrhopipi}) and (\ref{eq:inf Vrhoomegapi}). States in this appendix are normalized as $\braket{\boldsymbol{p}}{\boldsymbol{k}}=(2\pi)^3 \,2E_{\boldsymbol{p}}\,\delta^3(\boldsymbol{k}-\boldsymbol{p})$ unless specified other. 
Let $\ket{p_1^+ p_2^- p_3^0}$ and $T^\lambda(E;p_1,p_2,p_3)$ denote $\ket{\pi^+(\boldsymbol{p_1})\,\pi^-(\boldsymbol{p_2})\,\pi^0(\boldsymbol{p_3})}$ and T-matrix element $\bra{p_1^+ p_2^- p_3^0}T(E)\ket{\omega,\lambda}$ with $\lambda$ being the polarization of the $\omega$, respectively. Then,
\begin{widetext}
    \begin{align}
    T^\lambda(E;p_1,p_2,p_3) = \int \frac{d^3\boldsymbol{q}}{(2\pi)^6 \,2E_\rho(\boldsymbol{q})\,2E_\pi(\boldsymbol{q})}\sum\limits_{\sigma} 
    T^{I=0}_{\sigma}(E;p_1p_2p_3;q)\frac{1}{E-E_\rho(\boldsymbol{q})-E_\pi(\boldsymbol{q})} 
    V^{\lambda\sigma}_{\omega\to\rho\pi}(q)
    \end{align}
\end{widetext}
where $\sigma$ denotes the polarization of $\rho$ and
\begin{align}
    &T^{I=0}_{\sigma}(E;p_1p_2p_3;q) \equiv \bra{p_1^+p_2^-p_3^0}T^{I=0}(E)\ket{\rho(-\boldsymbol{q})\,\pi(\boldsymbol{q});\sigma}
    \\
    &V^{\lambda\sigma}_{\omega\to\rho\pi}(q) \equiv \bra{\rho(-\boldsymbol{q})\,\pi(\boldsymbol{q});\sigma}V^{I=0}\ket{\omega,\lambda}
\end{align}
With straightforward calculation it can be shown that
\begin{align*}
    T^\lambda(E;p_1,p_2,p_3) =  A^\lambda_{12}(E)  + A^\lambda_{23}(E) + A_{31}^\lambda(E)
\end{align*}
where $A_{ij}^\lambda(E)$ is short for $A^\lambda(E;p_i,p_j)$ given by
\begin{widetext}
    \begin{align}
        A^\lambda(E;p,k) = \sqrt{\frac{1}{6}} \sum\limits_{\sigma} \frac{1}{(2\pi)^3 2E_\rho(\boldsymbol{k})} V^\sigma_{\rho\to\pi\pi}(\boldsymbol{p}^*) \frac{W(E;\boldsymbol{k})-m_\rho^\B}{W(E;\boldsymbol{k})-m_\rho^\B-\Sigma(W(E;\boldsymbol{k}))} \frac{1}{E-E_\rho(\boldsymbol{k})-E_\pi(\boldsymbol{k})} V^{\lambda\sigma}_{\omega\to\rho\pi}(\boldsymbol{k})
    \end{align}
\end{widetext}
where $\sqrt{\frac{1}{6}}$ is the isospin factor, $\Sigma=\Sigma_{\pi\pi}+\Sigma_{\omega\pi}$ defined in Eq.(\ref{eq:self pipi},\ref{eq:self omegapi}), $\boldsymbol{p}^*$ is the momentum $\boldsymbol{p}$ boosted in $\rho$ rest frame since we have boosted the T-matrix element into $\rho$ rest frame and $W(E;\boldsymbol{k})\equiv \sqrt{\left(E-E_\pi(\boldsymbol{k})\right)^2 - \boldsymbol{k}^2}$. Therefore, the spin-averaged decay width is given by
\begin{align}
    &\bar{\Gamma} = \frac{(2\pi)^4}{(2\pi)^6 2m_\omega}\frac{1}{3} \sum\limits_{\lambda}\int \dd \Phi_3 \abs{T^\lambda(m_\omega;p_1,p_2,p_3)}^2
    \\
    &= \frac{(2\pi)^4}{(2\pi)^6 2m_\omega} \int \dd \Phi_3 \sum\limits_{\lambda} \left\{\abs{A_{23}^\lambda(m_\omega)}^2 \right. 
    \notag
    + 
    \\
    &  \qquad\qquad\qquad\qquad \left. 2\Re\left(A_{23}^\lambda(m_\omega) A_{31}^{\lambda*}(m_\omega)\right) \right\}
\end{align}
where $\dd\Phi_3$ is the Lorentz-invariant three-body phase space element defined in \cite{ParticleDataGroup:2022pth}. It is convenient to take $\dd\Phi_3 \propto \dd m_{12}\,\dd\Omega_3\,\dd\Omega_1^*$ and $\propto \dd m_{12}^2\,\dd m_{23}^2 \,\dd\Omega_{\text{Euler}}$ for the integration of $\abs{A}^2$ term and interference term, respectively. $V_{\rho\to\pi\pi}^\sigma$ and $V^{\lambda\sigma}_{\omega\to\rho\pi}$ can be written as two forms. The manifestly Lorentz-invariant one is convenient for the integration of THE interference term and the other for THE $\abs{A}^2$ term,
\begin{align}
       V^{\sigma}_{\rho\to\pi\pi}(\boldsymbol{k}) &= -(2\pi)^3 \, \sqrt{2} \, g_{\rho\pi\pi} \upepsilon_\mu(\boldsymbol{0},\sigma) \, (k_1^* - k_2^*)^\mu \,u_{\pi\pi}
       \\
       &= (2\pi)^\frac{9}{2}\left(\sqrt{2E_\pi(\boldsymbol{k}})\right)^2\sqrt{2m_\rho^\B} Y_{1\sigma}(\hat{\boldsymbol{k}}) V_{\pi\pi}(\boldsymbol{k})   
\end{align}
\begin{align}
     V^{\lambda\sigma}_{\omega\to\rho\pi}(\boldsymbol{k}) = (2\pi)^3 \, \sqrt{3} g_{\omega\rho\pi} \, \upepsilon^{\mu\nu\alpha\beta} \, P_\mu \,\upepsilon_\nu(\boldsymbol{0},\lambda) \,P_\alpha^\prime \,\upepsilon_{\beta}^*(-\boldsymbol{k},\sigma)
     \\
     = -\sqrt{8\pi}\,(2\pi)^3 \,m_\omega \,g_{\omega\rho\pi} \,C_{11}(1\lambda;\lambda-\sigma \sigma)\,Y^1_{\lambda-\sigma}(\hat{\boldsymbol{k}})  \abs{\boldsymbol{k}}
\end{align}
where $\upepsilon$ is the polarization vector, $u_{\pi\pi}$ and $V_{\pi\pi}$ are the form factor and potential defined in Eq.(\ref{eq:pipicutoff}) and Eq.(\ref{eq:inf Vrhopipi}), respectively. $p_{1/2}^*$ are the four-vectors of $\pi$, $P$ and $P^\prime$ are the four-vectors of $\omega$ and $\rho$, respectively. We do not introduce a form factor for $V_{\omega\to\rho\pi}$ since there is no loop integral related to it. 
Following that in Ref.\cite{Kleefeld:2001xd}, the interference term is given by,
\begin{widetext}
    \begin{align}
    &\sum\limits_{\lambda}\Re\left(A^\lambda_{23}(m_\omega)A_{31}^{\lambda*}(m_\omega)\right) = \frac{(2\pi)^6 g_{\rho\pi\pi}^2 g_{\omega\rho\pi}^2}{ 4 E_\rho(\boldsymbol{p_3}) E_\rho(\boldsymbol{p_1})}\frac{(m_{12}-m_\rho^\B)(m_{23}-m_\rho^\B)}{(m_\omega-E_\rho(\boldsymbol{p_3})-E_\pi(\boldsymbol{p_3})) (m_\omega-E_\rho(\boldsymbol{p_1})-E_\pi(\boldsymbol{p_1}))} u_{\pi\pi}(\boldsymbol{p_1}^*) u_{\pi\pi}(\boldsymbol{p_1}) \notag
    \\
    &\times \Re\left[\frac{1}{m_{12}-m_\rho^\B -\Sigma(m_{12})} \left(\frac{1}{m_{23}-m_\rho^\B -\Sigma(m_{23})}\right)^* \right]
    \begin{vmatrix}
        P^2 & P\cdot(p_2+p_3) & P\cdot (p_3 - p_2) \\
        P\cdot(p_1+p_2) & (p_1+p_2)\cdot(p_2+p_3) & (p_1+p_2)\cdot(p_3-p_2) \\
        P\cdot(p_2-p_1) & (p_2-p_1)\cdot(p_2+p_3) & (p_2-p_1)\cdot(p_3-p_2)
    \end{vmatrix} 
    \end{align}
\end{widetext}
$\abs{\boldsymbol{p_1}},\abs{\boldsymbol{p_3}}$ and the elements in the determinant can be easily expressed in terms of $m_{12}^2$ and $m_{23}^2$. On the other hand, the $\abs{A^\lambda}^2$ term is given by
\begin{widetext}
    \begin{align}
    \int \dd\Omega_3\dd\Omega_1^* \sum\limits_{\lambda} \abs{A^\lambda_{23}(m_\omega)}^2  =  \frac{m_\omega^2}{2}\left(\frac{1}{(2\pi)^3 2E_\rho(\boldsymbol{p_3})}\right)^2 \left(\frac{1}{m_\omega-E_\rho(\boldsymbol{p_3}) - E_\pi(\boldsymbol{p_3})}\right)^2 \abs{\frac{W(m_\omega;\boldsymbol{p_3})-m_\rho^\B}{W(m_\omega;\boldsymbol{p_3})-m_\rho^\B-\Sigma(W(m_\omega;\boldsymbol{p_3}))}}^2 \notag
    \\
    \times \left( (2\pi)^3 2\sqrt{2} g_{\rho\pi\pi} \abs{\boldsymbol{p_1}^*} u_{\pi\pi}(\abs{\boldsymbol{p_1}^*})  \right)^2 \left( (2\pi)^3 \sqrt{8\pi} m_\omega  g_{\omega\rho\pi} \abs{\boldsymbol{p_3}}   \right)^2 
    \end{align}
\end{widetext}
With these ingredients $\Gamma_{\omega\to\rho\pi\to3\pi}$ can then be calculated. 

In Sec.(\ref{sec:fitresult}) we found that when $V_{\omega\pi}=0$, $g_{\rho\pi\pi}$ and $\Lambda_{\rho\pi\pi}$ can be fixed at $7.07$ and $890$ MeV, respectively. It is expected that the introduction of the $\omega\rho\pi$ vertex would slightly shift these parameters, we try to take $g_{\rho\pi\pi}$ and $\Lambda_{\rho\pi\pi}$ at $7.40$ and $900$ MeV, respectively. Besides, we assume that $\Lambda_{\omega\rho\pi}=\Lambda_{\rho\pi\pi}$. The remaining two parameters $m_\rho^\B$ and $g_{\omega\rho\pi}$ are constrained by $\Gamma_{\omega\to3\pi}$ and $\delta^{\ell=1}_{\pi\pi\to\pi\pi}$. If we adopt $g_{\omega\rho\pi}=18$/GeV, which is close to that in Ref.\cite{Leinweber:2001ac}, and $m_\rho^\B=870$ MeV, it is found that
\begin{align}
    \Gamma_{\omega\to\rho\pi\to 3\pi} = 7.12 \,\text{MeV}
\end{align}
while the experimental value of the partial decay width $\Gamma^\text{ex}_{\omega\to3\pi} = 7.74(13)\,\text{MeV}$. The $10\%$ discrepancy is accounted for by the neglected direct interaction between the $\ket{\omega}$ and the $\ket{3\pi}$ channel\cite{Kleefeld:2001xd}. Furthermore, the phase shift defined by Eq.(\ref{eq:phaseshift}) can also be obtained and shown in Fig.$\,$\ref{fig:phase shift in the appendix}
The approximate consistence between the theoretical and experimental results justify the values adopted. Furthermore, it is interesting that the extrapolated $m_\rho^\B$ of HSC, MILC and Bulava \textit{et al}. in scheme C are nearly the same as that adopted here. 
~%
\begin{figure}[htbp]
    \centering
    \includegraphics[scale=0.45]{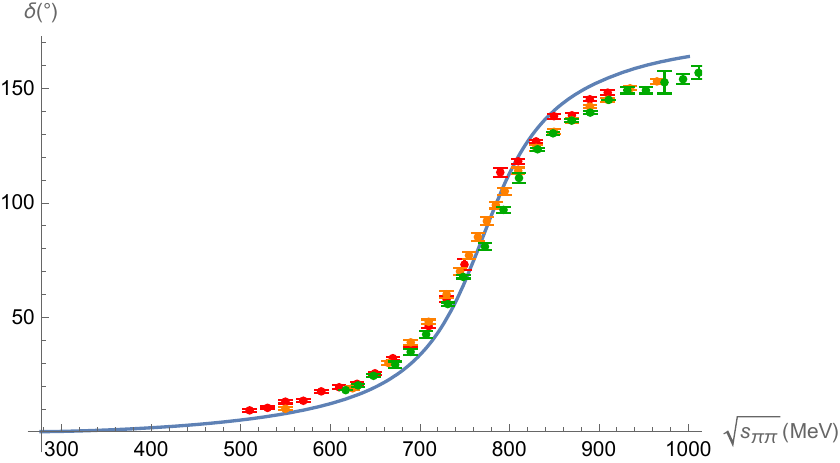}
    \caption{
    Phase shift calculation by Eq.(\ref{eq:phaseshift}) using parameters $g_{\omega\rho\pi}=18$/GeV, $\Lambda_{\omega\rho\pi}=\Lambda_{\rho\pi\pi}=900\,$MeV, $g_{\rho\pi\pi}=7.40$ and $m_\rho^\B=870\,$MeV at the physical pion mass $\mu_\pi$. The points in the figure are experimental values from Refs.~\cite{Protopopescu:1973sh, Estabrooks:1974vu, Hyams:1973zf}.
    }
    \label{fig:phase shift in the appendix}
\end{figure}

\section{Table of $C_{\Gamma\,,G}$}
\label{app:enstata}

In this appendix the values of $C_{\Gamma\,,G}$ are given. As in the main context, $C_{\Gamma\,,G}$ is the reduction coefficient. To be more specific, the non-zero space spanned by $\ket{\alpha;e_{\boldsymbol{n}},M}$ furnish a representation labelled by $\Gamma_\infty$ which is irreducible for $G_\infty$ but generally reducible for subgroup $G$, i.e., a restricted representation for $G$. Therefore, with $C_{\Gamma,G}$, the $\Gamma_\infty$ can be decompose into the direct sum of the irreducible representation of $G$. Furthermore, thanks to the Wigner-Eckart theorem, it is sufficient to take $a=1$ without loss of generality.
The non-vanishing $[C_{\Gamma\,,G}]_{M,a=1}$ relevant to the present work is given in the Table.(\ref{tab:my_label}). For more general result one can refer to, for example, Ref.~\cite{Li:2019qvh, Li:2021mob}.
\begin{table}[htbp]
    \centering
    \begin{tabular}{c@{\hspace{10pt}}c}
    \toprule
    {$(\Gamma,G)$} & {$\sum\limits_{M}[C_{\Gamma\,,G}]_{M,a=1}\ket{M}$} \\
    \midrule
    $(\mathrm{T_1},\mathrm{O_h})$ & $\frac{1}{\sqrt{2}}\ket{1} - \frac{1}{\sqrt{2}}\ket{-1}$
    \\ \midrule
    $(\mathrm{A_1},\mathrm{C_{4v}})$ & $\ket{0}$ 
    \\
    $(\mathrm{E},\mathrm{C_{4v}})$ & $\ket{-1}$
    \\ \midrule
    $(\mathrm{A_1},\mathrm{C_{3v}})$ & $\ket{0}$ 
    \\
    $(\mathrm{E},\mathrm{C_{3v}})$ & $\ket{-1}$
    \\\midrule
    $(\mathrm{A_1},\mathrm{C_{2v}})$ & $\ket{0}$
    \\
    $(\mathrm{B_1},\mathrm{C_{2v}})$ & $\frac{1}{\sqrt{2}}\ket{-1} + \frac{1}{\sqrt{2}}\ket{1}$
    \\
    $(\mathrm{B_2},\mathrm{C_{2v}})$ & $\frac{1}{\sqrt{2}}\ket{-1} - \frac{1}{\sqrt{2}}\ket{1}$
    \\ \bottomrule
    \end{tabular}
    \caption{Values of $[C_{\Gamma,G}]_{M,a=1}$ relevant to the present work. $\ket{M}$ is short for the state $\ket{\alpha;e_{\boldsymbol{n}}\,,M}$ defined in Eq.(\ref{eq:fin JMLS basis}) with $J=1$.}
    \label{tab:my_label}
\end{table}

\section{HEFT fits to lattice QCD spectra}
\label{app:fiting levels}
Here, we show all the other fit results for the various lattice QCD data sets collected besides that given in Fig.~\ref{fig: spectra example}.
%
\begin{figure*}[htbp]
    \centering    
  \subfloat[$m_\pi=176$ MeV]{\includegraphics[width=\textwidth]{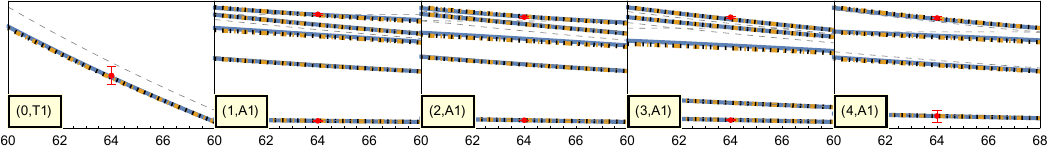}}\\
  \subfloat[$m_\pi=247$ MeV]{\includegraphics[width=\textwidth]{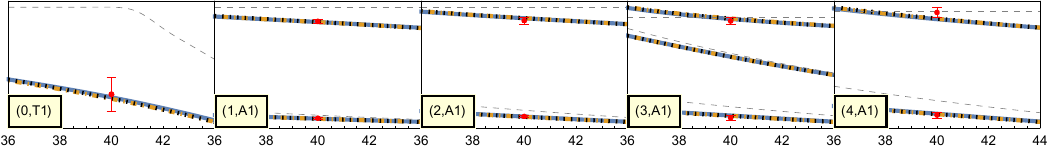}}\\
  \subfloat[$m_\pi=248$ MeV]{\includegraphics[width=\textwidth]{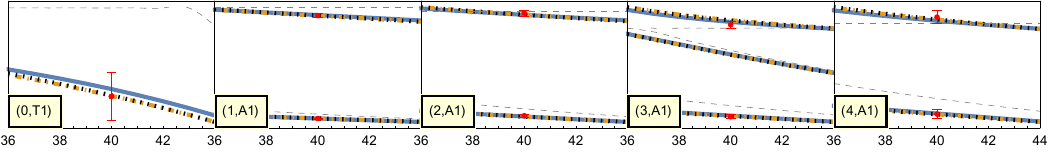}}\\
  \subfloat[$m_\pi=275$ MeV]{\includegraphics[width=\textwidth]{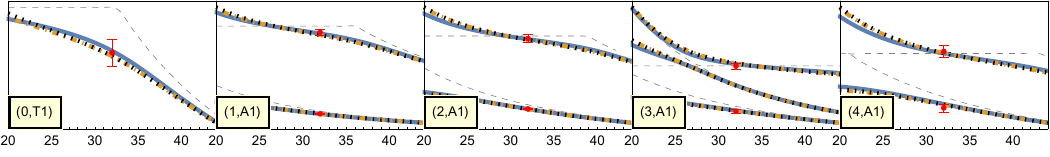}}\\
  \subfloat[$m_\pi=301$ MeV]{\includegraphics[width=\textwidth]{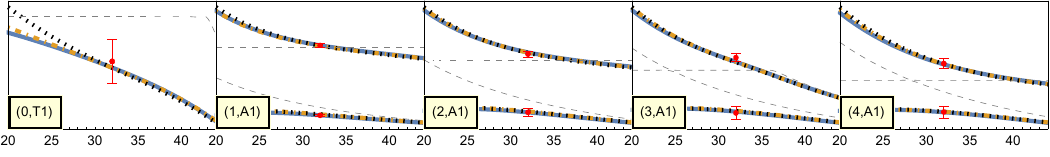}}\\
  \subfloat[$m_\pi=346$ MeV]{\includegraphics[width=0.9\textwidth]{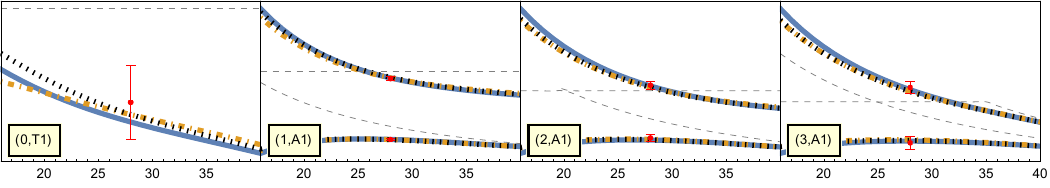}}\\
    \caption{ Same as in  Fig.~\ref{fig: spectra example} but for the MILC 
collaboration~\cite{Fu:2016itp}.}
\end{figure*}

\begin{figure*}[htbp]
    \centering
  \subfloat[$m_\pi=220$ MeV]{\includegraphics[width=\textwidth]{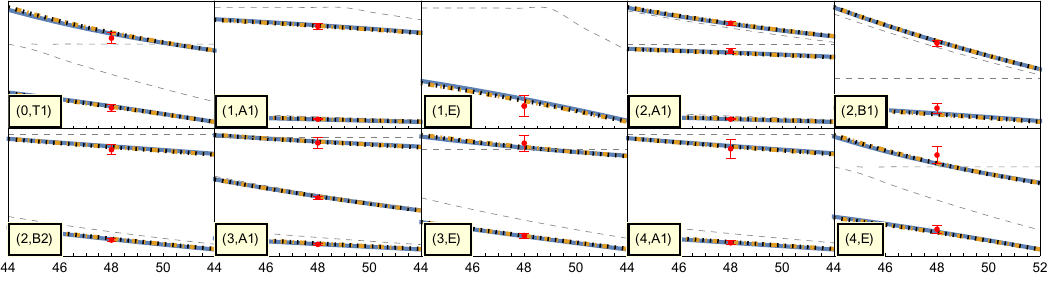}}\\
  \subfloat[$m_\pi=280$ MeV]{\includegraphics[width=\textwidth]{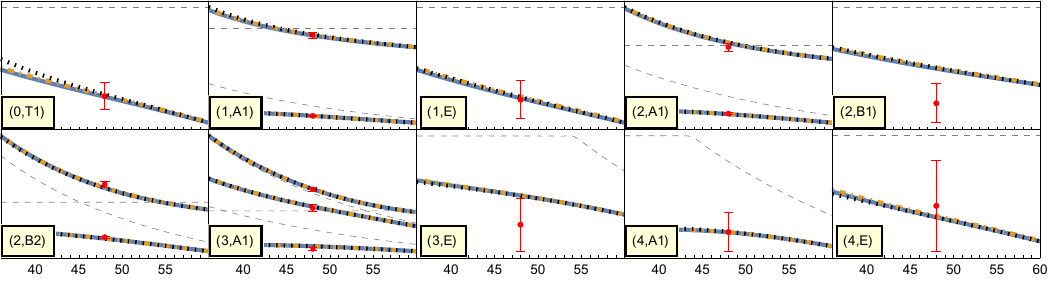}}\\
  \caption{Same as in Fig.~\ref{fig: spectra example} but for other $m_\pi$ by Bulava \textit{et al}.~\cite{Andersen:2018mau}.}
\end{figure*}
\begin{figure*}[htbp]
    \centering
    \subfloat[$m_\pi=316$ MeV]{\includegraphics[width=\textwidth]{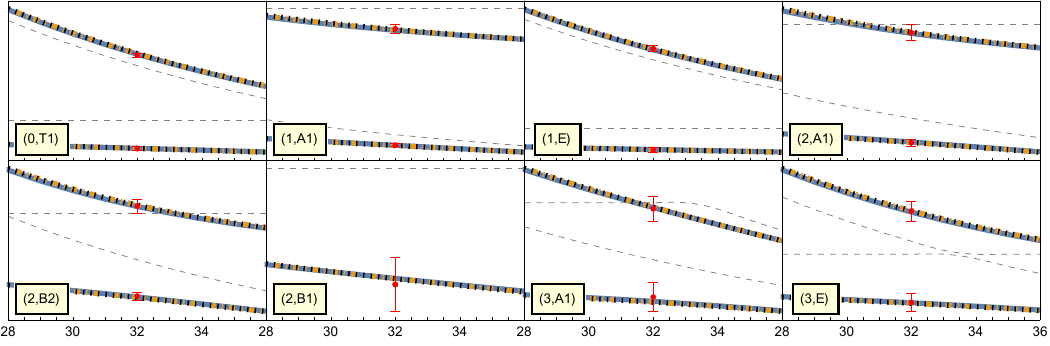}}
    \caption{
  Same as in Fig.~\ref{fig: spectra example} but for C.~Alexrandru \textit{et al}.~\cite{Alexandrou:2017mpi}
  }
\end{figure*}
\begin{figure*}[htbp]
    \centering
   \subfloat[$m_\pi=262$ MeV]{\includegraphics[width=0.8\textwidth]{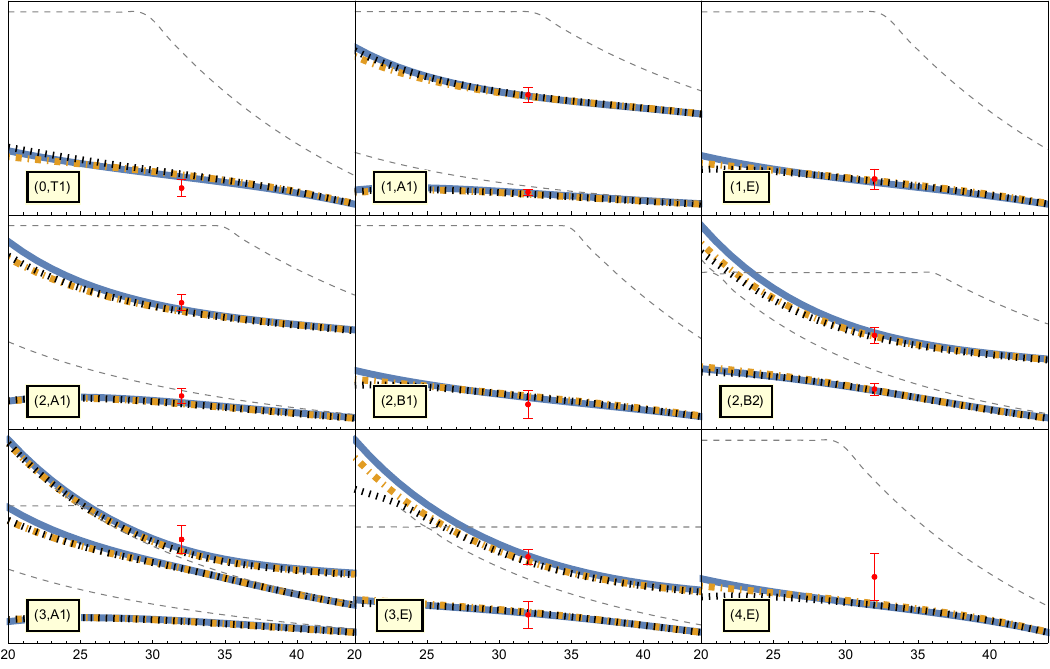}}\\
  \subfloat[$m_\pi=302$ MeV]{\includegraphics[width=\textwidth]{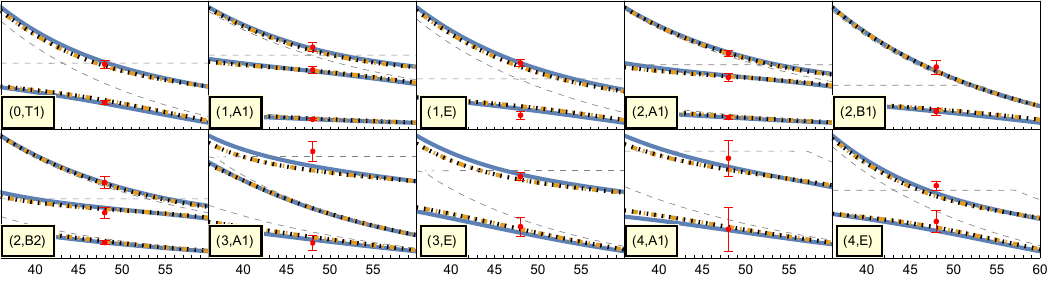}}\\
  \caption{Same as in Fig.~\ref{fig: spectra example} but for the ETMC collaboration~\cite{ExtendedTwistedMass:2019omo}.}
\end{figure*}
\begin{figure*}[htbp]
    \centering
  \subfloat[$m_\pi=322$ MeV]{\includegraphics[width=0.8\textwidth]{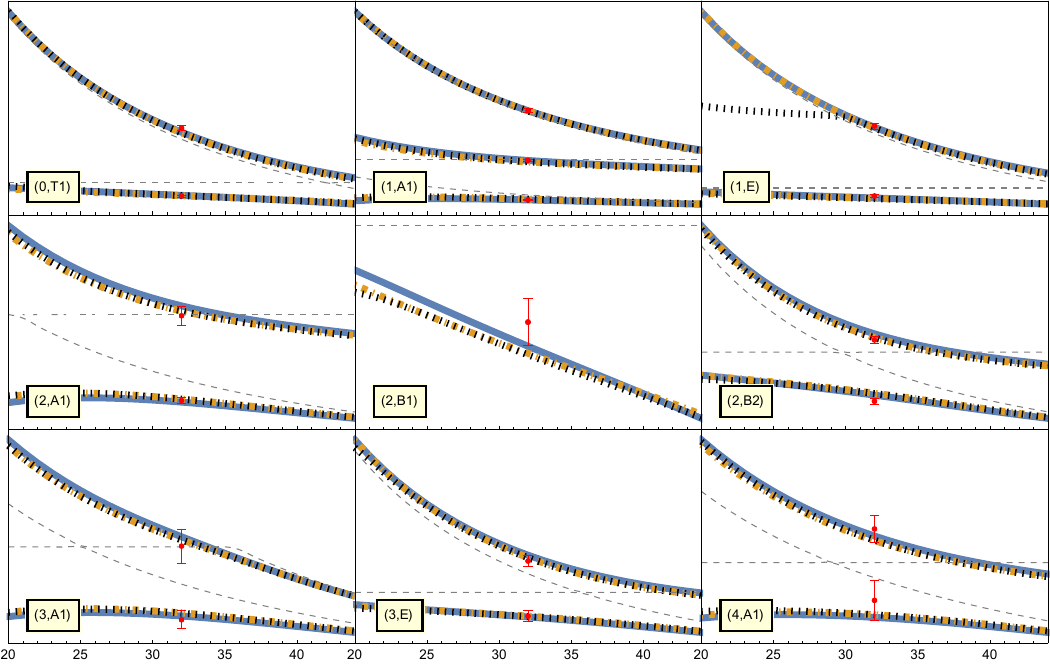}}\\
  \subfloat[$m_\pi=376$ MeV]{\includegraphics[width=\textwidth]{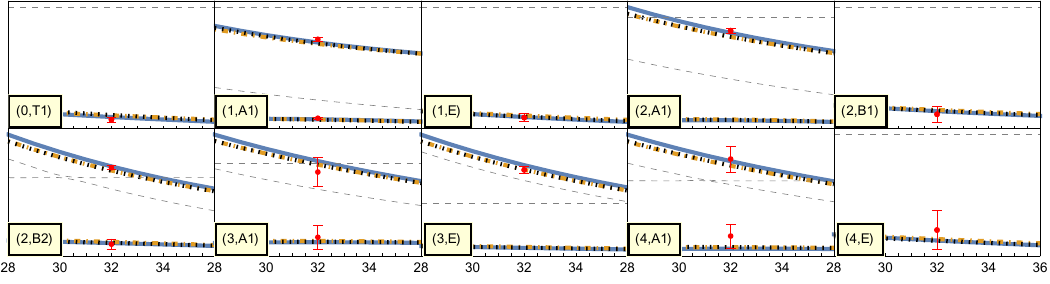}}\\
  \subfloat[$m_\pi=386$ MeV]{\includegraphics[width=\textwidth]{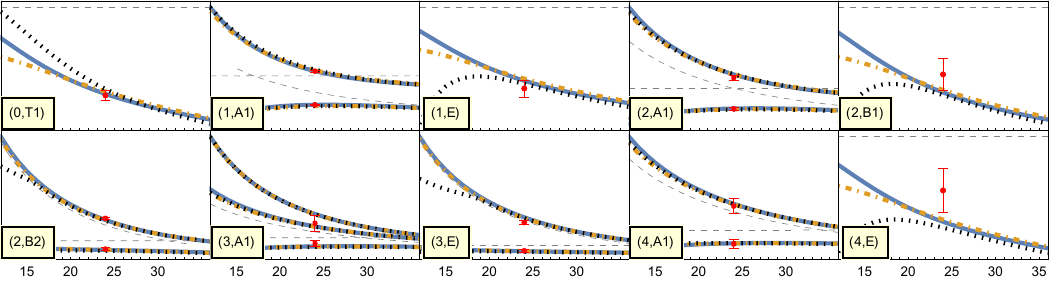}}\\
  \caption{Same as in Fig.~\ref{fig: spectra example} but for the ETMC collaboration~\cite{ExtendedTwistedMass:2019omo}. The turning point of the black dotted line for $m_\pi=322$ MeV is due to an avoided level crossing when the $\omega \, \pi$ channel is included. }
\end{figure*}
\begin{figure*}[htbp]
    \centering
  \subfloat[$m_\pi=236$ MeV]{\includegraphics[width=\textwidth]{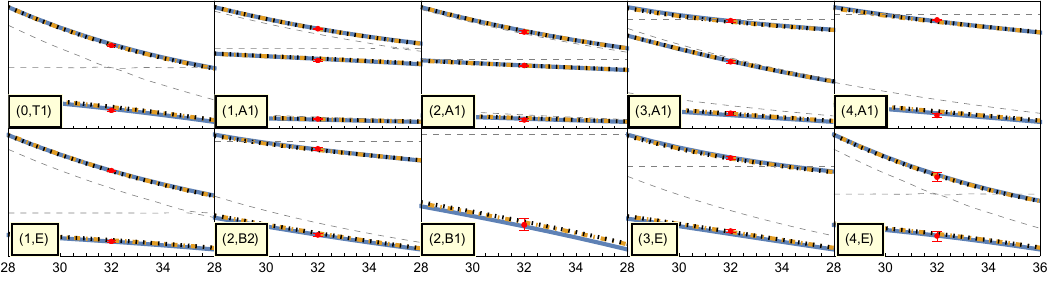}}\\
  \subfloat[$m_\pi=330$ MeV]{\includegraphics[width=\textwidth]{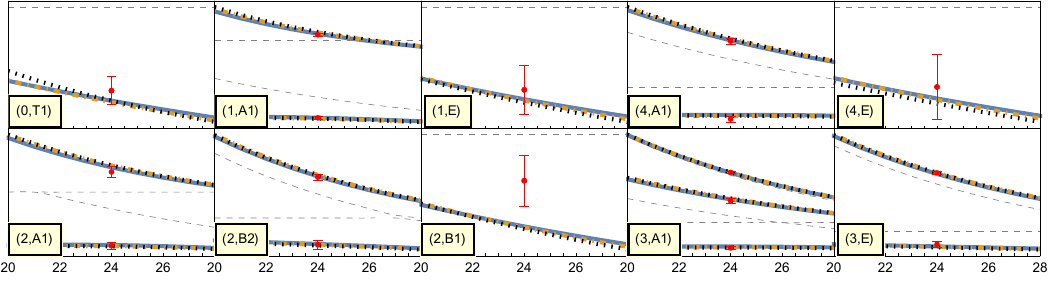}}\\
  \subfloat[$m_\pi=391$ MeV]{\includegraphics[width=0.8\textwidth]{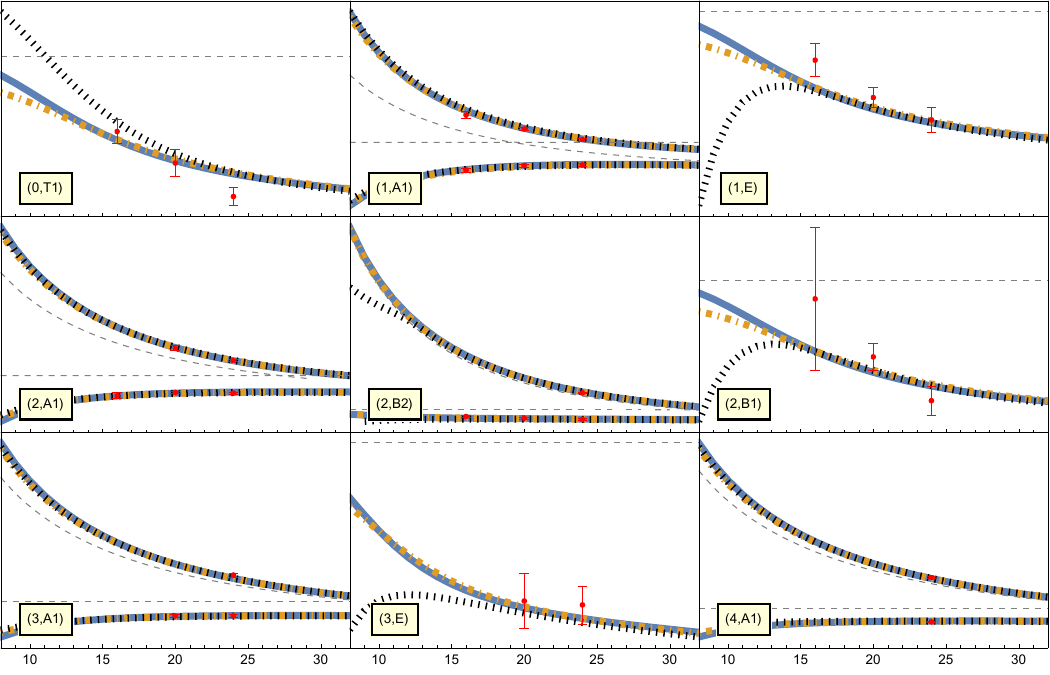}}\\
  \caption{Same as in Fig.~\ref{fig: spectra example} but for the HSC collaboration~\cite{Dudek:2012xn,Wilson:2015dqa,Rodas:2023gma}.}
\end{figure*}
\begin{figure*}
    \centering
     \includegraphics[width=0.9\textwidth]{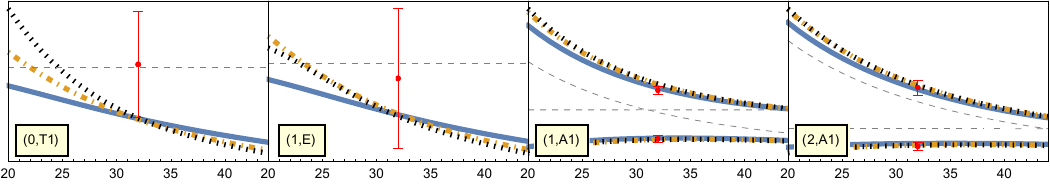}
     \caption{Same as in Fig.~\ref{fig: spectra example} but for the PACS-CS 
collaboration~\cite{Aoki:2020bew}.}
\end{figure*}

\begin{figure*}[htbp]
    \centering
  \subfloat[$m_\pi=$226 MeV]{\includegraphics[width=0.8\textwidth]{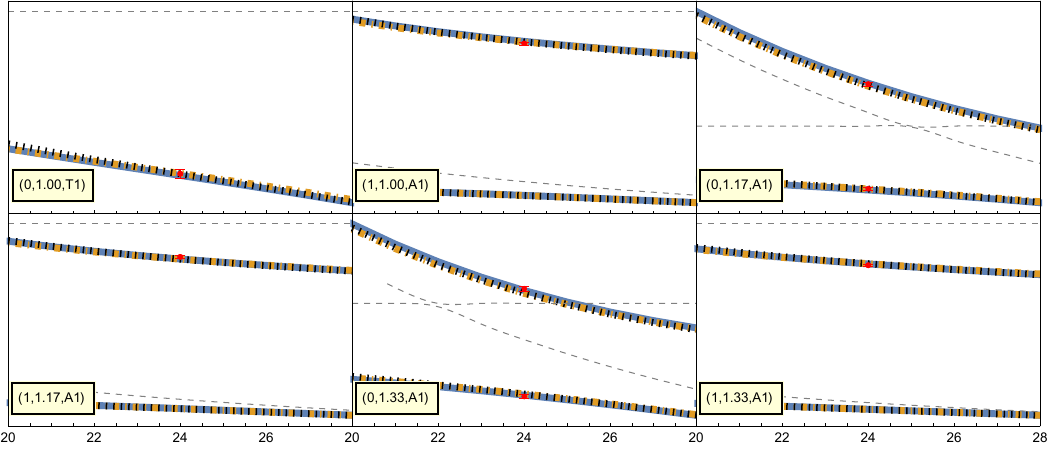}}\\
  \subfloat[$m_\pi=$315 MeV]{\includegraphics[width=0.8\textwidth]{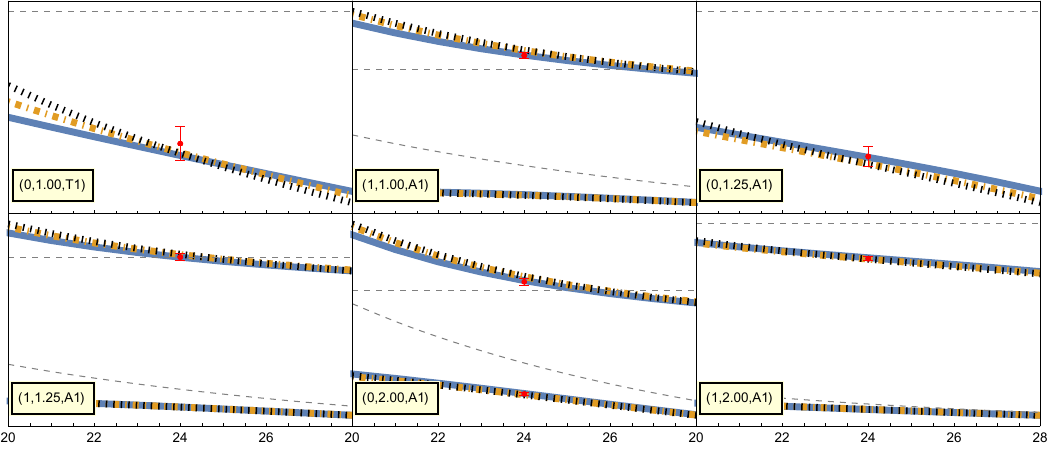}}\\
  \caption{Same as in Fig.~\ref{fig: spectra example} but for Guo \textit{et al}.~\cite{Guo:2016zos}.  Note that there is an additional quantity $\eta$ denoting the elongation factor in the yellow box compared to the others.}
  \label{fig:spectrum curve of L}
\end{figure*}

%
%

%


\end{document}